\documentclass[twocolumn]{aastex631}

\usepackage{rotating}

\begin{document}

\title{FEAST: JWST/NIRCam view of the Resolved Stellar Populations of the Interacting Dwarf Galaxies NGC~4485/NGC~4490}

\author[0009-0003-6182-8928]{Giacomo Bortolini}
\affiliation{Department of Astronomy, The Oskar Klein Centre, Stockholm University, AlbaNova, SE-10691 Stockholm, Sweden}

\author[0000-0001-6464-3257]{Matteo Correnti}
\affiliation{INAF Osservatorio Astronomico di Roma, Via Frascati 33, 00078, Monteporzio Catone, Rome, Italy}
\affiliation{ASI-Space Science Data Center, Via del Politecnico, I-00133, Rome, Italy}

\author[0000-0002-8192-8091]{Angela Adamo}
\affiliation{Department of Astronomy, The Oskar Klein Centre, Stockholm University, AlbaNova, SE-10691 Stockholm, Sweden}

\author[0000-0001-6291-6813]{Michele Cignoni}
\affiliation{Dipartimento di Fisica, Università di Pisa, Largo Bruno Pontecorvo 3, 56127, Pisa, Italy}
\affiliation{INAF - Osservatorio di Astrofisica e Scienza dello Spazio di Bologna, via Piero Gobetti 93/3, 40129 Bologna, Italy}
\affiliation{INFN, Largo B. Pontecorvo 3, 56127, Pisa, Italy}

\author[0000-0001-5618-0109]{Elena Sacchi}
\affiliation{Leibniz-Institut für Astrophysik Potsdam (AIP), An der Sternwarte 16, 14482 Potsdam, Germany}

\author[0000-0002-0986-4759]{Monica Tosi}
\affiliation{INAF - Osservatorio di Astrofisica e Scienza dello Spazio di Bologna, via Piero Gobetti 93/3, 40129 Bologna, Italy}

\author[0009-0003-6182-8928]{Göran Östlin}
\affiliation{Department of Astronomy, The Oskar Klein Centre, Stockholm University, AlbaNova, SE-10691 Stockholm, Sweden}

\author[0009-0009-2729-5025]{Anastasios Kapodistrias}
\affiliation{Department of Astronomy, The Oskar Klein Centre, Stockholm University, AlbaNova, SE-10691 Stockholm, Sweden}
\affiliation{Department of Teaching \& Learning, Stockholm University, Svante Arrhenius väg 20 A, 10691 Stockholm, Sweden}

\author[0000-0001-8068-0891]{Arjan Bik}
\affiliation{Department of Astronomy, The Oskar Klein Centre, Stockholm University, AlbaNova, SE-10691 Stockholm, Sweden}

\author[0000-0002-5189-8004]{Daniela Calzetti}
\affiliation{Department of Astronomy, University of Massachusetts, 710 North Pleasant Street, Amherst, MA 01003, USA}

\author[0000-0002-5259-4774]{Ana Duarte-Cabral}
\affiliation{Cardiff Hub for Astrophysics Research and Technology (CHART), School of Physics \& Astronomy, Cardiff University, The Parade, CF24 3AA Cardiff, UK}

\author[0000-0003-2442-6981]{Flavia Dell'Agli}
\affiliation{INAF Osservatorio Astronomico di Roma, Via Frascati 33, 00078, Monteporzio Catone, Rome, Italy}

\author[0000-0001-8608-0408]{John S. Gallagher}
\affiliation{Department of Astronomy, University of Wisconsin-Madison, 475 N. Charter Street, Madison, WI 53706, USA}

\author[0000-0003-4910-8939]{Benjamin Gregg}
\affiliation{Department of Astronomy, University of Massachusetts, 710 North Pleasant Street, Amherst, MA 01003, USA}

\author[0000-0002-3247-5321]{Kathryn~Grasha}
\altaffiliation{ARC DECRA Fellow}
\affiliation{Research School of Astronomy and Astrophysics, Australian National University, Canberra, ACT 2611, Australia}   
\affiliation{ARC Centre of Excellence for All Sky Astrophysics in 3 Dimensions (ASTRO 3D), Australia}

\author[0000-0002-0806-168X]{Thomas S.-Y. Lai}
\affiliation{IPAC, California Institute of Technology, 1200 E. California Blvd., Pasadena, CA 91125}

\author[0009-0009-5509-4706]{Drew Lapeer}
\affiliation{Department of Astronomy, University of Massachusetts, 710 North Pleasant Street, Amherst, MA 01003, USA}

\author[0000-0002-1000-6081]{Sean T. Linden}
\affiliation{Steward Observatory, University of Arizona, 933 N. Cherry Avenue, Tucson, AZ 85719, USA}

\author[0000-0003-1427-2456]{Matteo Messa}
\affiliation{INAF - Osservatorio di Astrofisica e Scienza dello Spazio di Bologna, via Piero Gobetti 93/3, 40129 Bologna, Italy}

\author[0000-0002-8222-8986]{Alex Pedrini}
\affiliation{Department of Astronomy, The Oskar Klein Centre, Stockholm University, AlbaNova, SE-10691 Stockholm, Sweden}

\author[0000-0003-2954-7643]{Elena Sabbi}
\affiliation{Space Telescope Science Institute, 3700 San Martin Drive, Baltimore, MD 21218, USA}
\affiliation{Gemini Observatory/NSF's NOIRLab, 950 North Cherry Avenue, Tucson, AZ 85719, USA}

\author[0000-0002-0806-168X]{Linda J. Smith}
\affiliation{Space Telescope Science Institute, 3700 San Martin Drive, Baltimore, MD 21218, USA}

\author[0000-0002-2199-0977]{Helena Faustino Vieira}
\affiliation{Department of Astronomy, The Oskar Klein Centre, Stockholm University, AlbaNova, SE-10691 Stockholm, Sweden}

\author[0000-0002-1821-7019]{John M. Cannon}
\affiliation{Macalester College, 1600 Grand Avenue, Saint Paul, MN 55105, USA}

\author[0000-0002-5542-1940]{Salvador Duarte Puertas}
\affiliation{Departamento de F\'{\i}sica Te\'orica y del Cosmos, Campus de Fuentenueva, Edificio Mecenas, Universidad de Granada, E-18071 Granada, Spain}

\author[0000-0003-2344-6593]{Carmelle Robert}
\affiliation{Universit\'e Laval \& Centre de recherche en astrophysique du Qu\'ebec, Qu\'ebec, QC, Canada}

\correspondingauthor{Giacomo Bortolini}
\email{giacomo.bortolini@astro.su.se}



\begin{abstract}
We present new JWST/NIRCam observations of the interacting dwarf galaxy system NGC~4485/NGC~4490 (a.k.a. Arp 269), obtained as part of the Cycle 1 Feedback in Emerging extrAgalactic Star clusTers (FEAST) program. NGC~4485 and NGC~4490 form the closest known pair of interacting late-type dwarf galaxies (at $\sim7.4$ Mpc), excluding the Magellanic Clouds. Near-infrared color-magnitude diagrams (CMDs) reveal a wide range of stellar populations in both galaxies, including young ($\lesssim 200$ Myr) upper main sequence stars, core helium-burning stars, and oxygen-rich asymptotic giant branch (AGB) stars. We also identify intermediate-age ($\sim 200$ Myr – $1$ Gyr) carbon-rich AGB stars and a well-populated old ($\gtrsim 1$ Gyr) red giant branch (RGB). The CMDs show two distinct bursts of star formation beginning $\sim30$ Myr and $\sim200$ Myr ago, the latter consistent with the most recent pericenter passage predicted by N-body simulations. The spatial distribution of stars reveals a tidal bridge extending from NGC~4485 and connecting to the disk of NGC~4490. Compact star-forming regions are seen along NGC~4490’s spiral arms, possibly originating from its infrared nucleus. A significant metallicity gradient is observed in the young stellar populations forming the bridge. These findings suggest that during the last pericenter passage, gas was stripped from NGC~4485 via tidal forces or ram pressure, accreted by NGC~4490, and mixed with in-situ material, fueling ongoing star formation. This system provides a unique nearby laboratory for studying how tidal interactions shape the star formation and chemical enrichment history of dwarf galaxies.
\end{abstract}

\keywords{Dwarf galaxies (416) --- Star formation (1569) --- Galaxy interactions (600) --- Galaxy stellar content (621)}


\section{Introduction} \label{sec:intro}


The $\Lambda$ cold dark matter (CDM) cosmological model is quite successful at reproducing most of the observed properties of our Universe \citep{White1978,Peebles1982,Frenk1988,White1991}, but still faces some observational challenges \citep{Tulin2018}. Among these, we highlight the prediction that present-day dwarf galaxies should be surrounded by a large population of satellites and stellar streams (e.g., \citealt{Diemand2008};\citealt{Wheeler2015}), a property not sufficiently verified by observations yet \citep{Kravtsov2004,Simon2007}. While the interaction of satellites with massive galaxies has been vastly confirmed by observations of spirals and giant ellipticals in the Local Volume \citep[e.g.,][]{Belokurov2006,Ibata2001,McConnachie2009,Martinez-Delgado2010,Crnojevic2016}, robust statistics on the satellites of dwarf galaxies and on their interactions is still missing \citep[e.g.,][and references therein]{Annibali2022}, except for a few interesting and well studied cases. For instance, there is clear evidence that the Magellanic Clouds have several satellites and interaction debris \citep[e.g.,][]{Belokurov2017,Kallivayalil2018,Pardy2020,Patel2020}, and that other late-type dwarfs have many tidal features or satellites \citep[e.g.,][]{Hancock2009,Bortolini2024,Zhang2024,Sacchi2024,Garner2025}, even when located in huge voids \citep[e.g., DDO~68, ][]{Izotov2009,Tikhonov2014,Annibali2016,Annibali2023,Correnti2025}. 

Studying interactions or accretions of satellite systems is not only a valuable testing ground for cosmological models, but is also key to understand how they impact the evolution of galaxies, affecting their morphology and kinematics, and providing mechanisms to trigger gas flows and starbursts \citep{Karachentsev2018}. In turn, understanding how exactly interactions and accretions affect the evolution of local galaxies is the only means to figure out how high-redshift and primordial galaxies may have evolved in the early Universe. 

Obtaining reliable statistics on the fraction of satellites and streams around dwarf galaxies is challenging because these systems, with their satellites and/or merger signatures, often have extremely low surface brightness and are difficult to detect. However, in the last decade an increasing effort is being invested in surveys specifically devoted to the search for small satellites around dwarf systems. Several ground-based, wide-field, photometric surveys are currently underway. the Solitary Local dwarfs survey (Solo), targeting 42 isolated dwarf galaxies within $3$ Mpc of the Milky Way \citep{Higgs2016}. The Magellanic Analog Dwarf Companions And Stellar Halos (MADCASH), designed to use resolved stars to map the virial volumes of galaxies within $\lesssim$ $4$ Mpc and with stellar masses of $1\mbox{--}7\times {10}^{9}\,\mathrm{{M}_{\odot}}$ \citep{Carlin2016}. The Smallest Scale Hierarchy (SSH) survey \citep{Annibali2020}, designed to determine the frequency and properties of interaction and merging events around a sample of 45 late-type dwarfs at distances between $\sim 1$ and $\sim 10$ Mpc, with the aim of revealing the presence of faint tidal features around them down to a surface brightness of $\mu_r \simeq 31$ mag/arcsec$^2$. The formation history and stellar populations of candidate satellites and streams can be studied in detail in the optical and near-infrared (NIR) thanks to the exceptional sensitivity and spatial resolution of space telescopes like the Hubble Space Telescope (HST) and the James Webb Space Telescope (JWST), which can resolve individual stars and star clusters in targets within approximately 20 Mpc (see \citealt{Tolstoy2009} and \citealt{Annibali2022} for in-depth reviews).

In this context, we present here the interesting case of the interacting system NGC~4485/NGC~4490 (a.k.a. Arp 269), located at an average distance of $\sim 7.4$ Mpc \citep{Calzetti2015,Sabbi2018}, one of the closest known double dwarf colliding systems  \citep{Pearson2018}. This pair of galaxies is one of the targets of the cycle 1 JWST program FEAST (Feedback in Emerging extrAgalactic Star clusTers, PI. A. Adamo). The galaxies are separated by only $\sim 7.5$ kpc in projection, with a prograde motion and a velocity separation of $\sim 30 \, \mathrm{km/s^{-1}}$ \citep{Pearson2018}. The closest massive galaxy to the pair is NGC~4369 ($\mathrm{M_{\star}} = 2.6 \, \times \, 10^{10} \, \mathrm{M_{\odot}}$), at a projected distance of $ 310$ kpc \citep{Pearson2018}. NGC~4490 is a low-luminosity late-type galaxy, imaged by the HST within the Treasury program LEGUS \citep{Calzetti2015} and classified there as SBd with a stellar mass of $\sim 2.5 \times 10^9 \, \mathrm{\mathrm{M_{\odot}}}$, a star formation rate (SFR) inferred from FUV of $2.8$ $\mathrm{\mathrm{M_{\odot}}/yr}$ \citep{Calzetti2015}, and an oxygen abundance derived from HII region emission lines of $\mathrm{12 + \log(O/H) = 8.37}$ \citep{Pilyugin2007}.  NGC~4485, also observed by LEGUS with HST, is classified as IBm, with a mass an order of magnitude lower than its companion ($\mathrm{M} \sim 3.2 \times 10^8 \, \mathrm{\mathrm{M_{\odot}}}$), a low recent SFR of $\sim 0.2$ $\mathrm{\mathrm{M_{\odot}}/yr}$, and no information on its chemical abundance. Interestingly, the stellar mass and metallicity of NGC~4490 are fairly similar to those of our closest late-type satellite, the Large Magellanic Cloud \citep{Bekki2005}, and the ratio of stellar masses of the two members ($\sim 8:1$, \citealt{Clemens1998}) is also similar to that of the two Magellanic Clouds. Moreover, HST imaging from the LEGUS survey confirmed the presence of a faint bridge that may connect the two galaxies, a structure first identified by \citet{ElmegreenD1998}. This feature corresponds to some bright star forming regions, as observed by \citet{Thronson1989} in H$\alpha$ and GALEX in the UV \citep{Smith2010}. HI and CO (1-0) emission in the bridge region has been detected using preliminary Very Large Array (VLA, Cannon et al., priv. comm) and IRAM 30 meter telescope observations (Linden, priv. comm), further suggesting the presence of dense gas in this extended tidal structure. We are therefore dealing with an intriguing analog of the Magellanic Cloud system \citep{Gardiner1996,Bekki2005,Besla2010,Guglielmo2014,Pardy2018}, with the important difference that our targets are not affected by the presence of a more massive companion like the Milky Way. 

Despite the wealth of observations, the system's interaction history remains elusive. Some authors propose that NGC 4490 is a relatively young galaxy, approximately 2 Gyr old, undergoing continuous star formation at a steady rate for at least the past $100$ Myr \citep{ElmegreenD1998,Clemens1999,Clemens2000}. In this scenario, the two galaxies would have interacted only once, with their closest approach occurring around $100$ Myr ago. In contrast, recent N-body and test-particle simulations by \cite{Pearson2018} suggest that the galaxies have experienced two encounters, with perigalactic passages taking place approximately $1.3$ Gyr and $230$ Myr ago. To complicate the picture even more, \citet{Lawrence2020} recently found that NGC~4490 has a double nucleus morphology, with one nucleus visible in the optical and one only in the infrared (IR, see Figure 3 of \citealt{Lawrence2020}). Due to their mass and luminosity being comparable to other nuclei found in interacting galaxy pairs, the authors argued that NGC~4490 might be itself a merger remnant, which is now further interacting with NGC~4485. Finally, recent observations with the Five-hundred-meter Aperture Spherical radio Telescope (FAST) revealed an impressive HI tidal tail extending approximately 100 kpc both to the south and north \citep{Liu2023}, indicating a highly complex and extended dynamical environment \citep{Karachentsev2024}.

In this paper, we focus on the properties of the resolved stellar populations of NGC 4485, NGC 4490, and their bridge, based on our deep NIRCam imaging and analysis of NIR color-magnitude diagrams (CMDs). Our goal is to shed light on the complex interaction history of this system. A more detailed study of its star formation history, based on the synthetic CMD method, is currently in preparation (Bortolini et al., in prep). The paper is organized in three main Sections. In Section \ref{sec:datareduction} we describe in detail our JWST data, reduction pipeline, point spread function (PSF) fitting photometry and artificial stars test routine. In Section \ref{sec:colormagnitudediagrams} we present our analysis of NGC 4485/NGC 4490 CMDs, the properties of their different stellar populations, and their spatial distributions within the system. Finally, in Section \ref{sec:summary and conclusions} we summarize our results.


\begin{figure}[thbp!]
    \centering
    \includegraphics[width=0.5\textwidth]{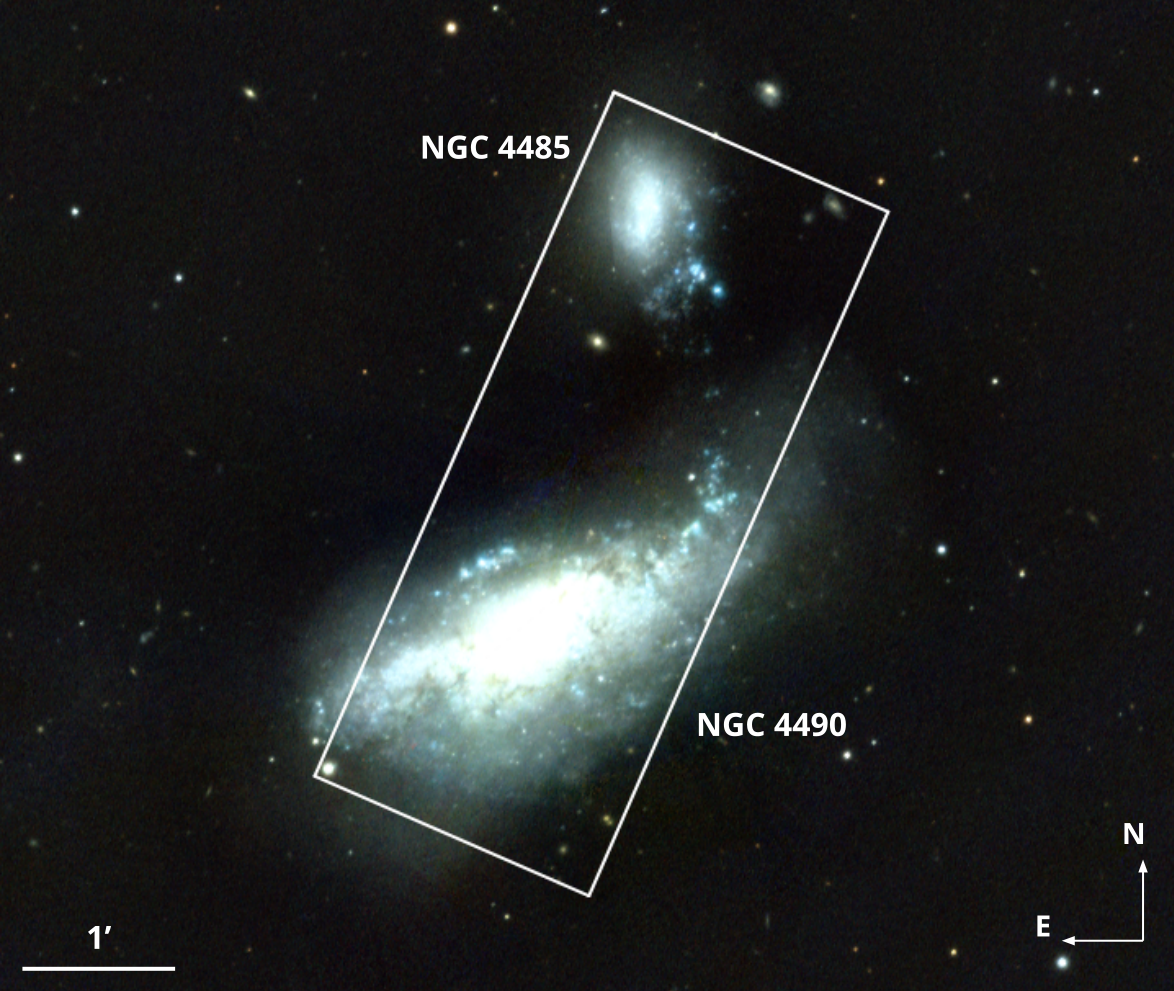}
    \caption{Footprint for the FEAST JWST/NIRCam mosaic of NGC~4485/NGC~4490 (white rectangle) overlaid upon an optical-band image adopted from the Panoramic Survey Telescope and Rapid Response System (Pan-STARRS, \citealt{Chambers2016}). The two dwarf galaxies are labeled. North is up, east is to the left.}
    \label{fig:1}
\end{figure}

\section{Observations \& Data Reduction} \label{sec:datareduction}
We present JWST observations of the NGC~4485/NGC~4490 system, carried out as part of the Cycle 1 program FEAST (PI: A. Adamo). In particular, the target was imaged by NIRCam \citep{Rieke2023} with a combination of narrow, medium, and wide band filters (i.e., F187N, F115W, F150W, F200W, F300M, F335M, F405N, and F444W), and by MIRI \citep{Rieke2015} in the F560W and F770W filters. The NIRCam mosaics (footprint shown in Figure \ref{fig:1}) have been observed with a FULLBOX 4TIGHT primary dither pattern, along with 2 small dithers resulting in 8 single exposures covering $6' \times 2.2'$ when combined. For the purpose of this study, we focus solely on the NIRCam's F115W and F200W filters. Among all the near-infrared short wavelengths filters available, this combination offers the best trade-off between high resolution and depth, allowing us to probe and resolve the galaxys' stellar populations, even in the crowded fields of their central regions. We note that the F200W filter includes the Pa$\alpha$ recombination line. Nevetheless its presence has negligible effect on the stellar photometry. We refer the reader to Adamo et. al. (in prep) for an in-depth overview of the FEAST survey. 

\begin{figure*}[thbp!]
    \centering
    \includegraphics[width=1\textwidth]{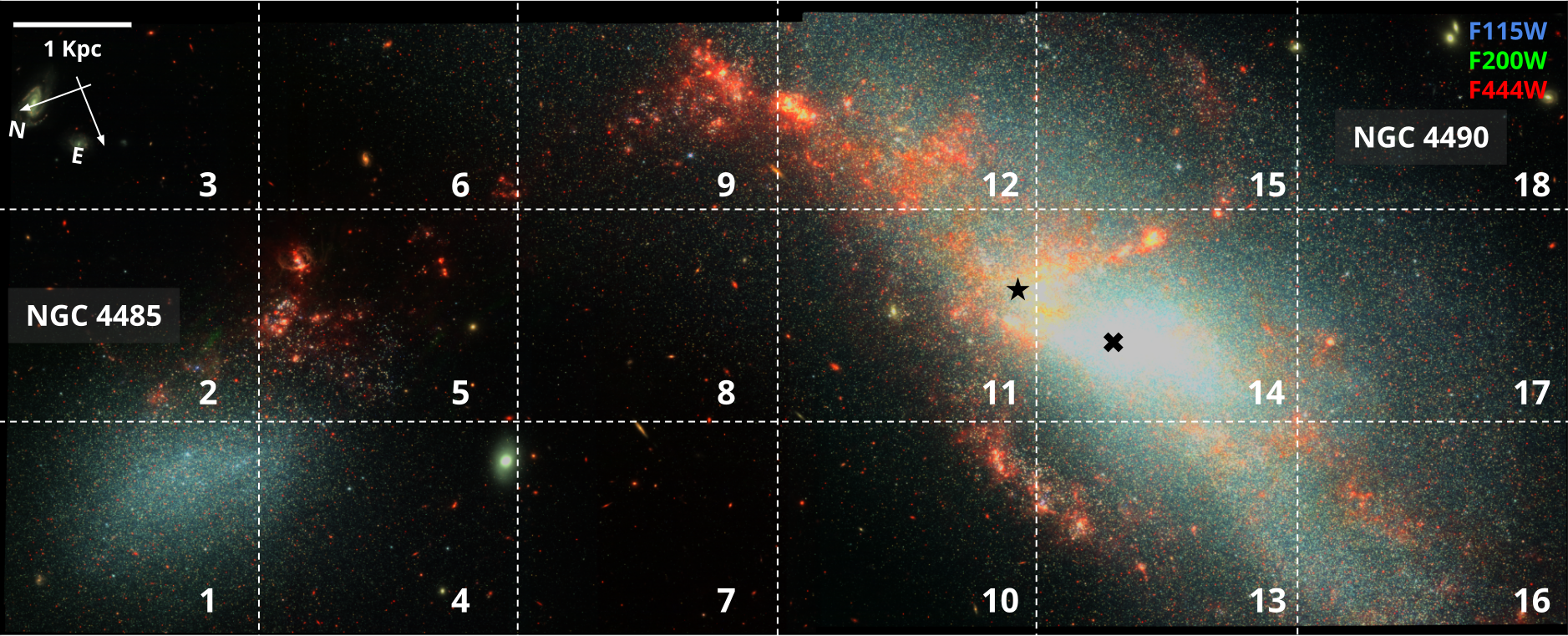}
    \caption{Color composite image of NGC~4485/NGC~4490, constructed by combining the NIRCam observations obtained as part of the FEAST program in F115W (blue channel), F200W (green channel), and F444W (red channel). The field of view is divided into 18 tiles, which are outlined with dashed white lines and numbered from left to right, bottom to top. The F115W and F200W trace the stellar continuum, while dust emission from star-forming regions shines in the F444W. The two dwarf galaxies are labeled. The black cross and star symbols mark the position of the optical and infrared nuclei \citep{Lawrence2020}. Particularly bright background galaxies can also be identified in tile 3, 4, 6, and 18}.
    \label{fig:2}
\end{figure*}

Figure \ref{fig:2} shows a color composite image of the NIRCam mosaic of NGC~4485/NGC~4490, with the F115W and F200W filters in blue and green tracing the stellar continuum, and the F444W filter in red tracing the dust re-emission in the NIR of UV photons from massive stars in young star-forming regions. This emission follows the spiral arms, as seen in optical images, and seems to be centered on the IR nucleus (black star, \citealt{Lawrence2020}). We also report the presence of a series of red and bright HII regions, forming a striking bridge from the southwest to the northeast side of the image, connecting the two dwarfs main bodies dominated by the blue stellar continuum.
\par 

\subsection{Stellar Photometry} \label{sec:stellarphotometry}
We performed PSF fitting photometry on both F115W and F200W images simultaneously, using the latest version of the software package \texttt{DOLPHOT 2.0} \citep{Dolphin2000,Dolphin2016} NIRCam module \citep{Weisz2024}. We used the F200W drizzled {\it *i2d.fits} image, aligned to GAIA DR3 \citep{GAia2023} as an astrometric reference, while the actual photometry was performed on the individual {\it *crf.fits} images. The \textit{*.fits} files have been processed using the pipeline versioning information CAL VER = 1.11.3, CRDS VER = 11.16.20, and CRDS CTX = jwst 1100.pmap. We ran the photometry following the prescriptions suggested in \citet{Weisz2024}. The final magnitudes are calibrated in the Vegamag system, using updated zero points aligned with the recommended Sirius-Vega-based calibration system \citep{Weisz2024}.

Due to the large field of view covered by our observations and the extremely high number of measured point sources, we ran \texttt{DOLPHOT} on 18 similarly sized portions of the reference image (hereafter referred to as \textit{tiles}; see Figure \ref{fig:2}) to speed up the reduction process, and the artificial star tests (see Section \ref{sec:artificialstartest}). Each of the 18 `raw' photometric catalogs were inspected and culled to remove as many interlopers as possible (i.e., background galaxies, blends, spurious detections) while retaining the largest number of bona-fide stars. To do so, we exploit several diagnostic parameters given by \texttt{DOLPHOT}\footnote{See \texttt{DOLPHOT} manual at \url{http://americano.dolphinsim.
com/dolphot/} for an in depth description of the output parameters.}. We first required {\tt Object\_Type} $<$ 1, {\tt quality\_flag} $\leq$2, {\tt SNR} $>$ 2. We then applied further `bell-shaped' selections cuts on the parameters {\tt sharpness} and {\tt crowding}, using the same prescriptions described in \citet{Correnti2025b}. The {\tt sharpness} selection was based on the distribution of {\tt sharpness} versus magnitude. We included sources with $|${\tt sharpness}$|$ $<$ 0.075 or those lying within $\pm 2\sigma$ of the local mean. For the {\tt crowding} parameter, we selected sources with {\tt crowding} $<$ 0.1, or alternatively, those within $3\sigma$ of the local mean.

\subsection{Artificial star tests} \label{sec:artificialstartest}
Artificial star tests (ASTs) are essential in photometric studies of crowded fields, providing the most robust and consistent method for estimating uncertainties and completeness as a function of a star's magnitude, color, and position in the image \citep{Cignoni2016,Weisz2024}. Taking advantage of our tiled photometry approach, we injected 1 million artificial stars in each tile (for a total of 18 million artificial stars), spatially distributed to mimic the light-profile of the image, and cover uniformly the magnitude range $(16,30)$ and color range $(-2,4)$ of our data (see CMDs in Figure \ref{fig:4}). Each artificial star was injected separately in each tile, in order not to bias our completeness estimation. Photometry was then carried out using the same \texttt{DOLPHOT} parameters as in the original run, and the output catalogs filtered using the same selection cuts described in Section \ref{sec:stellarphotometry}. An input artificial star that is detected and that passes the quality cuts is considered a recovered star. We can estimate the photometric errors of a star as a function of its color and magnitude from the distribution of the difference between output and input magnitudes, while the completeness can be evaluated taking the ratios between the number of recovered and injected artificial stars.

\begin{figure*}[thbp!]
    \centering
    \includegraphics[width=1\textwidth]{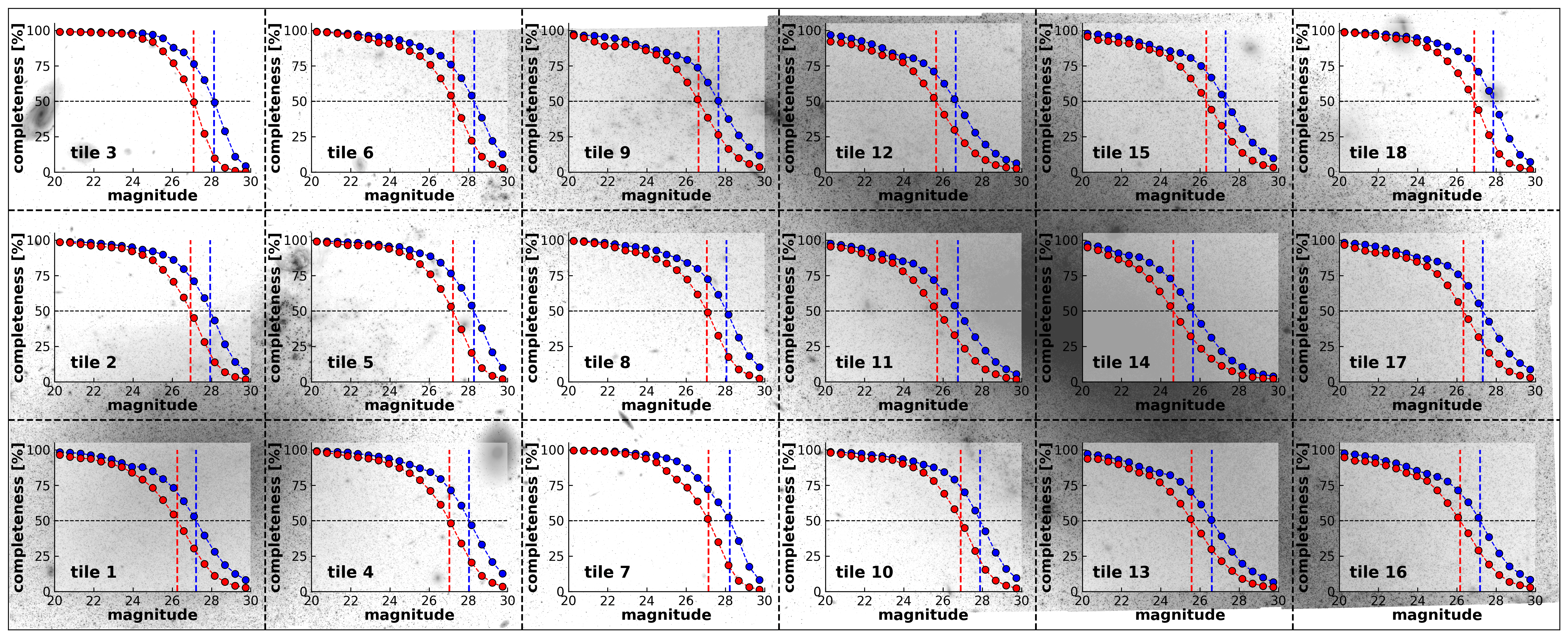}
    \caption{NIRCam's F200W image, with superimposed for each tile the completeness of the photometry as a function of magnitude, derived from our ASTs, in both the F115W (blue) and F200W (red) filters. The $50$\% completeness level is marked by the dashed-gray line, while the magnitude at which this limit is reached in each filter is marked by the dashed-vertical lines.}
    \label{fig:3}
\end{figure*}

Figure \ref{fig:3} shows the color-averaged photometric completeness as a function of input magnitude and position in the image (i.e., for each tile), in both the F115W (blue dots) and F200W (red dots) filters. The completeness levels are displayed on top of the F200W reference image. In each subplot, the $50$\% completeness level is marked by a dashed-gray line, while the magnitude at which it is reached in each filter is marked by the dashed-vertical lines. The $50$\% completeness level is reached around $\mathrm{m_{F115W}} = 27$ mag and $\mathrm{m_{F200W}} = 26$ mag in the central part of NGC~4485, while it is $\sim 1$ mag shallower in the more crowded center of NGC~4490. 

\section{Resolved Stellar population analysis} \label{sec:colormagnitudediagrams}

\subsection{Color-Magnitude Diagrams} \label{sec:Color-Magnitude Diagrams}

\begin{figure*}[thbp!]
    \centering
    \includegraphics[width=1\textwidth]{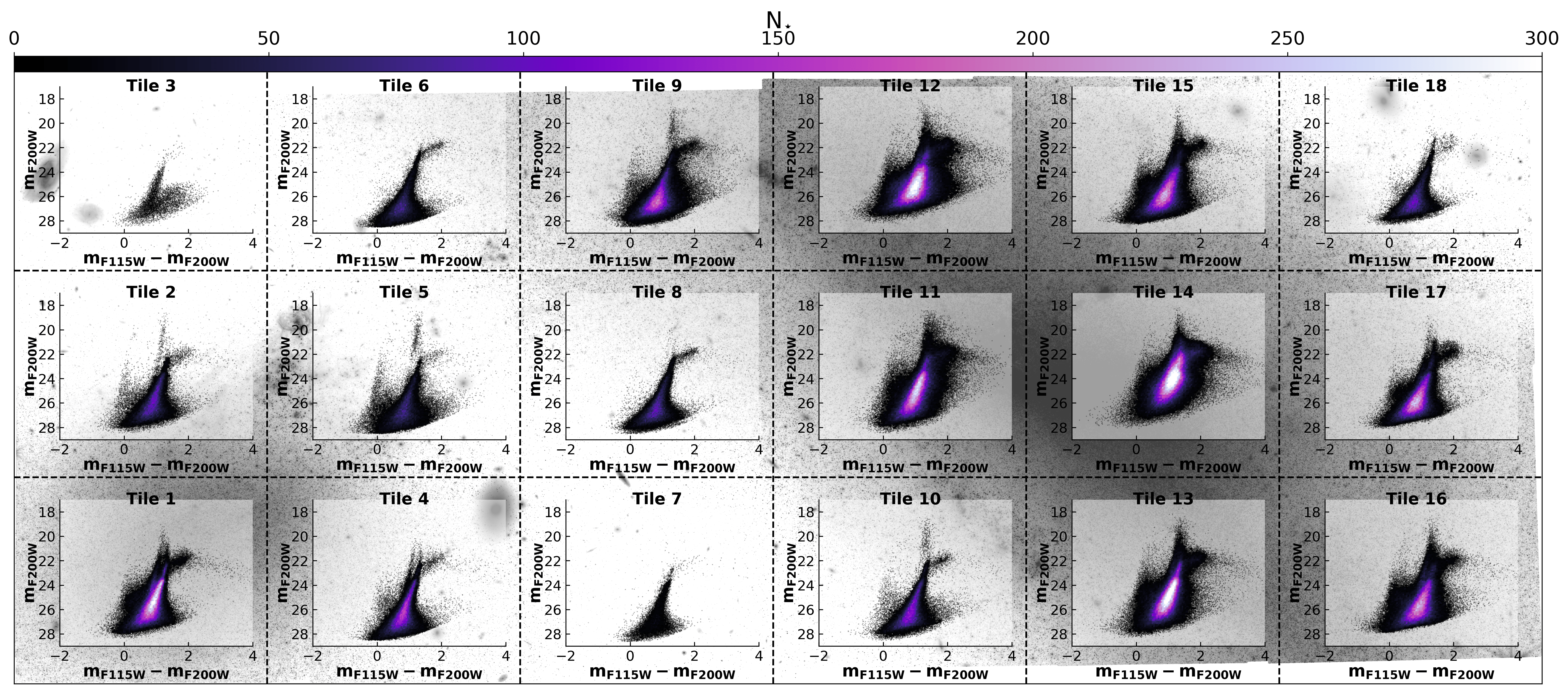}
    \caption{$\mathrm{m_{F200W}}$ vs. $\mathrm{m_{F115W}}-\mathrm{m_{F200W}}$ CMDs for each tile, displayed on top of the F200W reference image of the NGC~4485/NGC~4490 system. The high-density regions of the CMDs have been binned and color coded according to the number of stars in each bin (see colorbar on the top).}
    \label{fig:4}
\end{figure*}

Figure \ref{fig:4} shows the $\mathrm{m_{F200W}}$ vs. $\mathrm{m_{F115W}}-\mathrm{m_{F200W}}$ CMDs over-plotted on the position of the respective tiles, after we apply the quality cuts presented in Section \ref{sec:stellarphotometry}. To help the reader's eye in exploring the different stellar populations present in the diagrams, the high-density regions have been binned and color-coded based on stellar density. At first glance, a wide range of stellar populations are immediately visible, with the CMDs exhibiting different shapes and sizes across the two galaxies and even from tile to tile. The main features, in most of the tiles, include a bright-blue vertical sequence at $\mathrm{m_{F115W}}-\mathrm{m_{F200W}} \sim 0$ mag composed of young upper main-sequence (MS) stars and more evolved core He-burning stars at the blue edge of their loops, i.e. the so-called blue loop (BL) phase. Some examples can be seen in tile 1, 2, 5, 9, 10, 12, 15, 16. A distinct red vertical sequence is present at $\mathrm{m_{F115W}}-\mathrm{m_{F200W}} \sim 1$ mag and $\mathrm{m_{F200W}} \lesssim 23$ mag, composed of massive ($\mathrm{8 \, \mathrm{M_{\odot}} \lesssim M \lesssim 25 \, \mathrm{M_{\odot}}}$) core He-burning stars, but this time on the red edge of their loops, also known as the red supergiant (RSG) phase. Strong examples of this phase appear to be in tile 2, 5, 9, and 10, which are tiles that clearly show young HII regions based on the JWST imaging in Figure \ref{fig:2}. This stellar evolutionary phase occurs in the life of bright-massive stars and is characterized by a strong anti-correlation between the star age and its luminosity (i.e., the luminosity decreases with increasing age, e.g., \citealt{Maeder2001}, \citealt{Bressan2012}). All these properties make these stars particularly well suited to be used as `astronomical clocks' in distant galaxies \citep[see among others][]{Dohm-Palmer1997,Sacchi2016}. In between these two sequences, we observe a sparse cloud of young He-burning stars, in the process of going from the red to the blue edge (and vice versa) of their loops. A red `finger', running almost parallel to the RSG sequence around $\mathrm{m_{F115W}}-\mathrm{m_{F200W}}\sim 1.25$ mag and $\mathrm{m_{F200W}} \sim 21$ mag, is also present in some tiles (e.g., tile 1 and 4). This region of the CMD is mainly populated by intermediate-mass ($4 \, \mathrm{M_{\odot}} \lesssim M \lesssim 8 \, \mathrm{M_{\odot}}$) oxygen-rich asymptotic giant branch (O-rich AGB) stars. A distinct group resembling a red `tail' is present at $\mathrm{m_{F115W}}-\mathrm{m_{F200W}} \sim 2$ mag and $\mathrm{m_{F200W}} \sim 22$ mag, primarily consisting of less massive ($1.2 \, \mathrm{M_{\odot}} \lesssim M \lesssim 4 \, \mathrm{M_{\odot}}$) AGB stars in their thermally pulsing phase (TPAGB). During this phase, inner-shell convective episodes, known as the third dredge-up, transport a substantial amount of carbon to the stellar surface. This carbon is then released into the interstellar medium through strong winds that produce mainly carbon dust \citep{Nanni2013,Ventura2014}, causing these stars to appear redder compared to their O-rich counterparts. As a result, these stars are also referred to as carbon-rich (C-rich) AGB stars. Last but not least, low-mass ($M \lesssim 1.0 \, \mathrm{M_{\odot}}$) red giant branch (RGB) stars dominate the entire region of the diagrams around $\mathrm{m_{F115W}}-\mathrm{m_{F200W}} \sim 1$ mag, going from our faintest detection limit at $\mathrm{m_{F200W}}\sim 28$ mag, up to their tip located at $\mathrm{m_{F200W}} \sim 23.8$ mag.

\begin{figure*}[thbp!]
    \centering
    \includegraphics[width=0.95\textwidth]{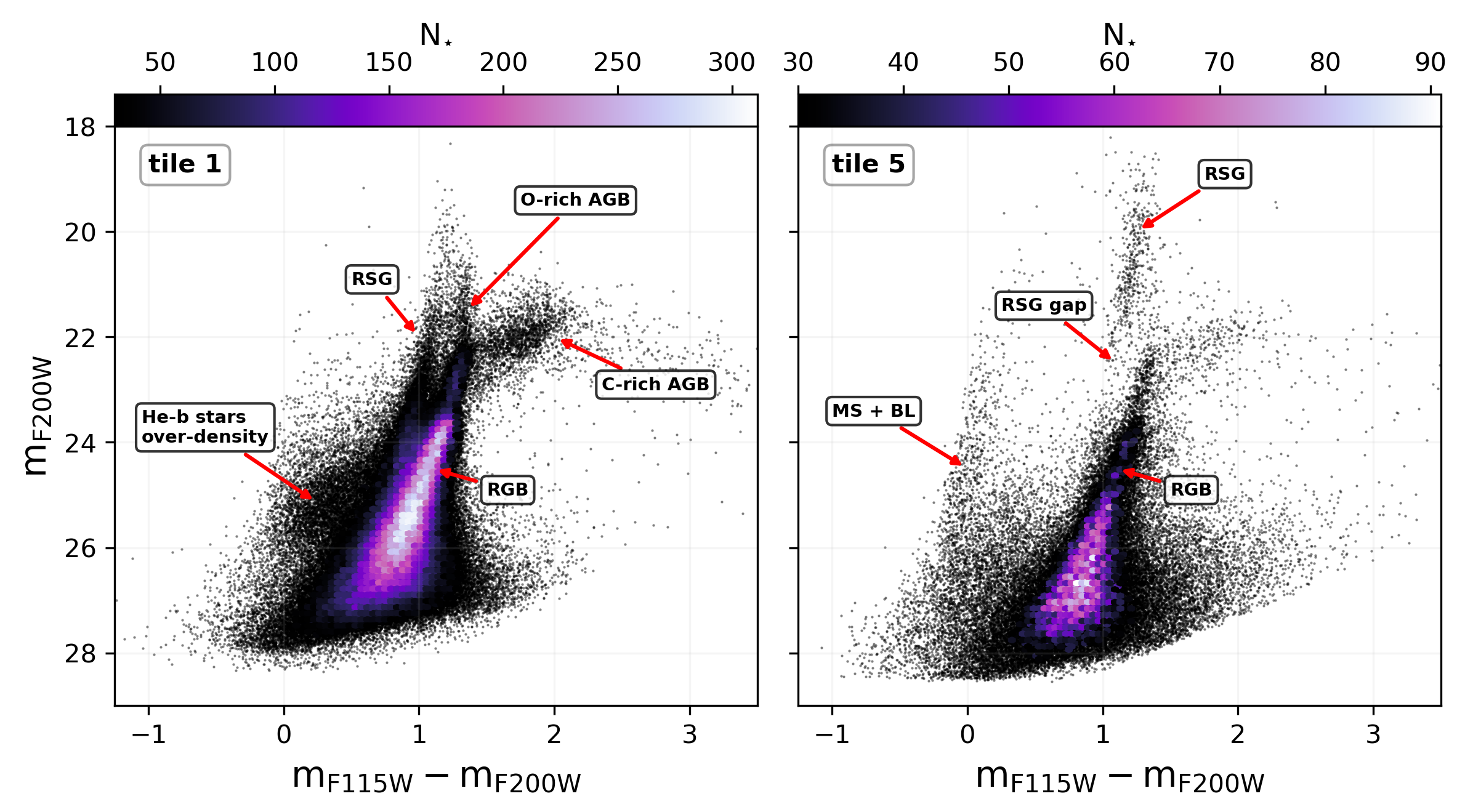}
    \caption{Zoom-ins of the CMDs of tile 1 (left panel) and tile 5 (right panel), with some of the features discussed in Section \ref{sec:Color-Magnitude Diagrams} \& \ref{sec:Isochrones comparison} highlighted and labeled accordingly. The high-density regions in the diagrams are binned and color-coded, ranging from black to white according to the number of stars in each bin.}
    \label{fig:5}
\end{figure*}

Figure \ref{fig:5} presents zoom-ins on the CMDs of tile 1 (left panel) and tile 5 (right panel), with some of the main features discussed above highlighted and labeled. In tile 1, these include the O-rich AGB `finger', the C-rich AGB red `tail', the RGB sequence, and an over-density of blue He-burning stars. In tile 5, a gap in the RSG sequence is marked, together with the MS plus BL stars blue sequence. These and other features will be analyzed in more detail in the next section, through comparisons with PARSEC-COLIBRI stellar isochrones.

\subsection{Isochrones comparison} \label{sec:Isochrones comparison}
\begin{figure*}[thbp!]
    \centering
    \includegraphics[width=1\textwidth]{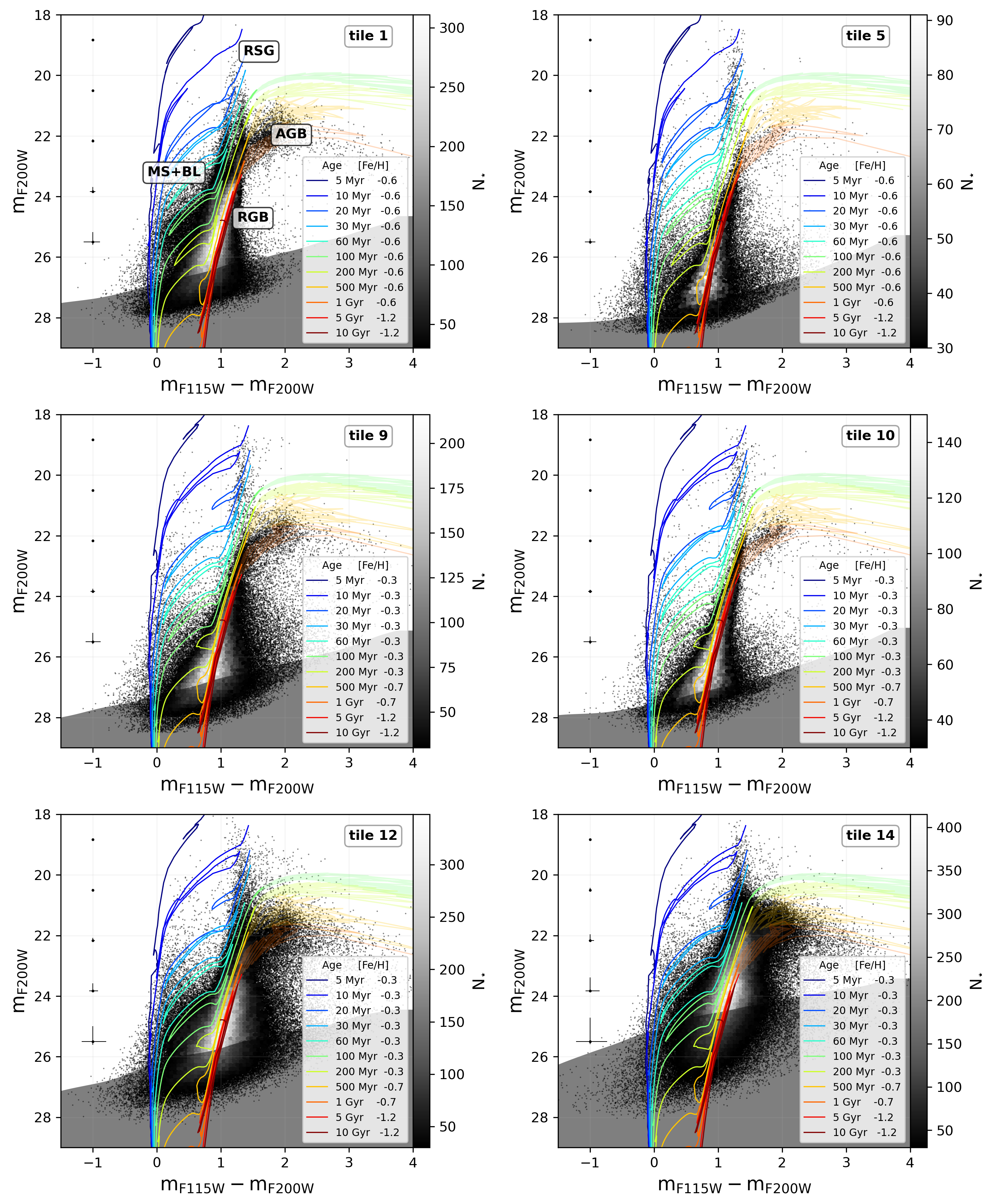}
    \caption{CMDs of tile 1, 5, 9, 10, 12, and 14 covering different regions of both NGC~4485 and NGC~4490. Each CMD is binned and color coded according to the density of stars. Different sets of PARSEC-COLIBRI isochrones are also plotted, assuming a distance modulus of $\mathrm{(m-M)_{0}} = 29.70$ and a reddening of $\mathrm{E(B-V)} = 0.15$, color-coded according to age and metallicity (see legend on the bottom right of each panel). The TPAGB phase is shown with a lower alpha value. Photometric errors estimated from ASTs are shown on the left-hand side. The shaded area marks the $50\%$ completeness limit. Some of the stellar evolutionary phases mentioned in the text are labeled in tile 1 CMD.}
    \label{fig:6}
\end{figure*}
Figure \ref{fig:6} shows the CMDs for tiles 1 and 5, which cover the main body of NGC 4485 and part of the bridge, respectively, as well as tiles 9, 10, 12, and 14, which sample the remaining portion of the bridge and regions of particular interest within NGC 4490. To more easily identify the different stellar populations present, we plotted a set of PARSEC-COLIBRI isochrones \citep{Bressan2012,Marigo2017,Pastorelli2020}, assuming a distance modulus of $\mathrm{(m-M)_{0}} = 29.70$ \citep{Calzetti2015,Sabbi2018}  and a color excess of $\mathrm{E(B-V)} = 0.15$ \citep{Sabbi2018}. We could not appreciate any substantial difference in the distance of the two dwarfs in our study, but an in depth analysis using the RGB-tip method \cite[e.g.,][and references therin]{Lee1993,Salaris1998,Bellazzini2004,Makarov2006,Bellazzini2024} is beyond the scope of this paper. The isochrones are color-coded according to their age and metallicity (see legend on the right-hand side of each panel). The TPAGB phase in the isochrones is shown with a lower opacity, due to the intrinsic uncertainties affecting the modeling of this particular stellar phase \citep{Ventura2018}. Error bars on the left-hand side of the plots are the photometric errors as a function of magnitude, estimated via the extensive ASTs (see Section \ref{sec:artificialstartest}). Finally, the shaded area marks the portion of the CMDs that lie below the $50\%$ completeness level (i.e. every source in the region has a $50\%$ or higher chance of not being recovered by our photometry). Unless otherwise specified, from this point onward we will refer only to stars above the $50\%$ completeness limit in their respective tiles. To fit the color of the very young stars in the RSG sequences, we assumed two different metallicity values for the two galaxies: $\mathrm{[Fe/H]} = -0.6$ for NGC~4485 (i.e. tile 1 and 5); and $\mathrm{[Fe/H]} = -0.3$ for the rest of the tiles belonging to NGC~4490 (see Section \ref{sec:metallicitygradient} for a more in depth discussion). On the other hand, an old population of stars with $\mathrm{[Fe/H]} = -1.2$ seems to well fit the color of the RGB branch of both NGC~4485 and NGC~4490. 
\par

The CMDs of tiles 1 and 5 show an extremely tight and well-populated RSG sequence at $\mathrm{m_{F115W}} - \mathrm{m_{F200W}} \sim 1$ mag, covering a magnitude range from $\mathrm{m_{F200W}} \sim 18$ mag to $\sim 24$ mag. In tile 5 (i.e. in the bridge), the RSG sequence displays a clear gap in the magnitude interval $24$ mag $\lesssim \mathrm{m_{F200W}} \lesssim 22$ mag (see also Figure \ref{fig:5}). The most likely explanation for such a paucity of stars in this region of the CMD is that the galaxy experienced a period of low star formation activity between $\sim 50 - 200$ Myr ago (as indicated by the isochrones' ages), followed by a strong burst that began around $30$ Myr ago and lasted until very recent epochs ($\sim 5$ Myr ago). Another very interesting feature, present in tile 1 (i.e, in the main body of NGC~4485) is a second red `finger', above the RGB tip, running almost parallel to the RSG sequence (also marked in Figure \ref{fig:5}). This sequence is mostly populated by relatively massive O-rich AGB stars with ages that span between $\sim 100$ and $\sim 200$ Myr. Less evolved He-burning stars of the same age form also a striking over-density cloud around $\mathrm{m_{F115W}}-\mathrm{m_{F200W}} \sim 0.5$ mag and $\mathrm{m_{F200W}} \sim 25$ mag (already highlighted in Figure \ref{fig:5}). The definitive proof that these stars trace the same population is given by the fact that in tile 5, where we do not observe a prevalent O-rich AGB finger, the fainter BL over-density is also absent. The tight magnitude range covered by the He-burning stars over-density, allows us to infer the birth epoch of this population with great precision around $100 \mbox{--} 200$ Myr ago (see Isochrones in tile 1 panel of Figure \ref{fig:5}). Moreover, the lower number of stars brighter or fainter than this feature points to the bursty nature of this star formation event. Interestingly, the age of this burst seems to agree remarkably well with the time of the second close encounter between the two galaxies, independently predicted by \citet{Pearson2018} N-body simulations. In the CMDs of both tiles we recovered a red and bright ($\mathrm{m_{F200W}} \sim 22$ mag) horizontal `tail', with a large color range, from $\sim 1.5$ mag to $\sim 4$ mag, populated by C-rich AGB stars with ages between $\sim 200$ Myr and $\sim 1$ Gyr. Finally, the most populated feature in both CMDs is the RGB, composed of very old (age $\gtrsim 1$ Gyr), low-mass stars that trace the earlier (possibly the earliest) epochs in the life of the galaxy.

The CMD of tile 9 in Figure \ref{fig:6}, covering NGC~4490's side of the bridge region, displays many similarities with that of tile 5, including a similar paucity of stars in the RSG sequence around $\mathrm{m_{F200W}} \sim 22.5$ mag, pointing to a synchronized onset of this starburst event. Interestingly, tile 10's CMD seems to display a young population with similar characteristics to those observed in the bridge (tile 9), an RSG sequence with a clear break and the absence of an O-rich AGB finger, even though tile 10 covers a bright star-forming region to the northeast of the bridge. The projected distance between these two star-forming regions ($\sim 2$ kpc) limits the possible explanations for this tightly synchronized burst to the gravitational interaction with NGC~4485. Finally, tiles 12 (covering a bright star forming region next to the NGC 4490 part of the bridge) and 14 (which covers the optical center of NGC~4490; \citealt{Lawrence2020}) show all of the main features discussed above. These include a very bright RSG sequence without any breaks, a well-populated BL sequence corresponding to isochrones with ages between approximately $100$ and $200$ Myr, and bright O-rich and C-rich AGBs. A pronounced RGB sequence is also present. The CMDs of both tiles, which cover intrinsically crowded regions of the galaxy, are affected by shallower and more uncertain photometry due to the more intense crowding conditions. Readers interested in a detailed analysis of the remaining tiles' CMDs are referred to Appendix \ref{app:Analysis of the remaining CMDs}.

\subsection{Metallicity Gradient} \label{sec:metallicitygradient}
\begin{figure}[thbp!]
    \centering
    \includegraphics[width=0.45\textwidth]{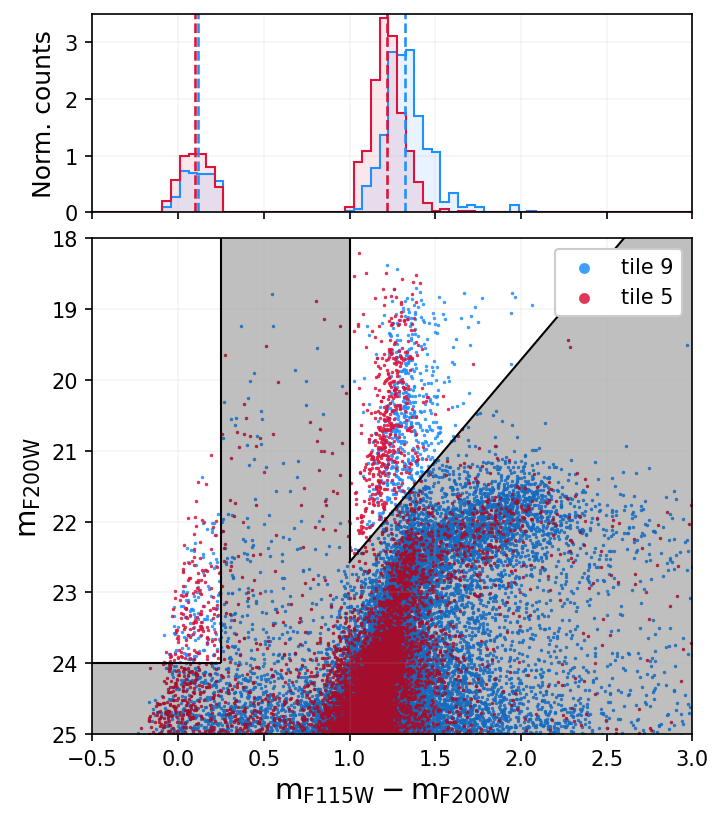}
    \caption{Bottom panel: Comparison between the CMDs of tile 5 (red points) and tile 9 (blue points). Top panel: Normilized distributions as a function of color of the upper-MS and RSG sequences for tile 5 and 9, after the exclusion of the shaded area to avoid contamination from other stellar evolutionary phases and photometric errors. The dashed vertical lines represent the median colors of tile 5 and tile 9 upper-MS and RSG sequences.}
    \label{fig:7}
\end{figure}

Aside from different stellar chemical compositions, we also explored other potential explanations for the substantial shift in color of the young RSG sequence stars across the bridge (see Section \ref{sec:Isochrones comparison}), such as photometric errors or internal reddening. The former mechanism can be rejected, since it would cause only a higher spread around the mean color of the RSG sequence rather than a systematic shift in color, while the latter would cause also a corresponding shift in the blue-edge of the upper main-sequence stars, which we do not observe. For instance, Figure \ref{fig:7} shows the comparison between the CMDs of tile 5 and 9 (see bottom panel), focusing in particular on the difference in the average color between the two RSG sequences and the blue edge of the upper-MS. We observe a clear shift of $\sim 0.2$ mag in the median color of the two RSG sequences (vertical dashed red and blue line respectively), without seeing a corresponding shift in the color of the upper-MS (see top panel of Figure \ref{fig:7}). This finding confirms the presence of a significant chemical composition gradient among the young stellar populations of the bridge between the two dwarf galaxies. Recently, Duarte Puertas et al. (in prep) mapped the system's gas-phase oxygen abundance as part of the SIGNALS survey \citep{Rousseau-Nepton2019} with the imaging-Fourier transform spectrograph SITELLE \citep{Drissen2019}. Interestingly, the authors found that the gas in NGC 4485 is substantially more metal-poor than that in NGC 4490, with a clear gradient across the tidal bridge connecting the two galaxies, consistent with what we find in the young stellar component. Moreover they found a pocket of metal-poor gas corresponding to the star forming complex on the northeast side of NGC 4490 (located in tile 10 and 13).

Taken together, our results and those of Duarte Puertas et al., (in prep) point to a scenario in which the most recent pericenter passage (which occurred $\sim 200$ Myr ago) stripped metal-poor gas from NGC 4485, likely due to a combination of tidal interactions and ram pressure \citep{Mayer2006}. Subsequently, this gas appears to have partially mixed with the more metal-rich gas of NGC 4490, fueling simultaneously the very recent burst ($\lesssim 30$ Myr) of star formation in the bridge (i.e. tile 5 and 9) and in the northwest side of NGC 4490 (i.e. tile 10).

\subsection{The Spatial Distribution of Stars}
\begin{figure}[thbp!]
    \centering
    \includegraphics[width=0.5\textwidth]{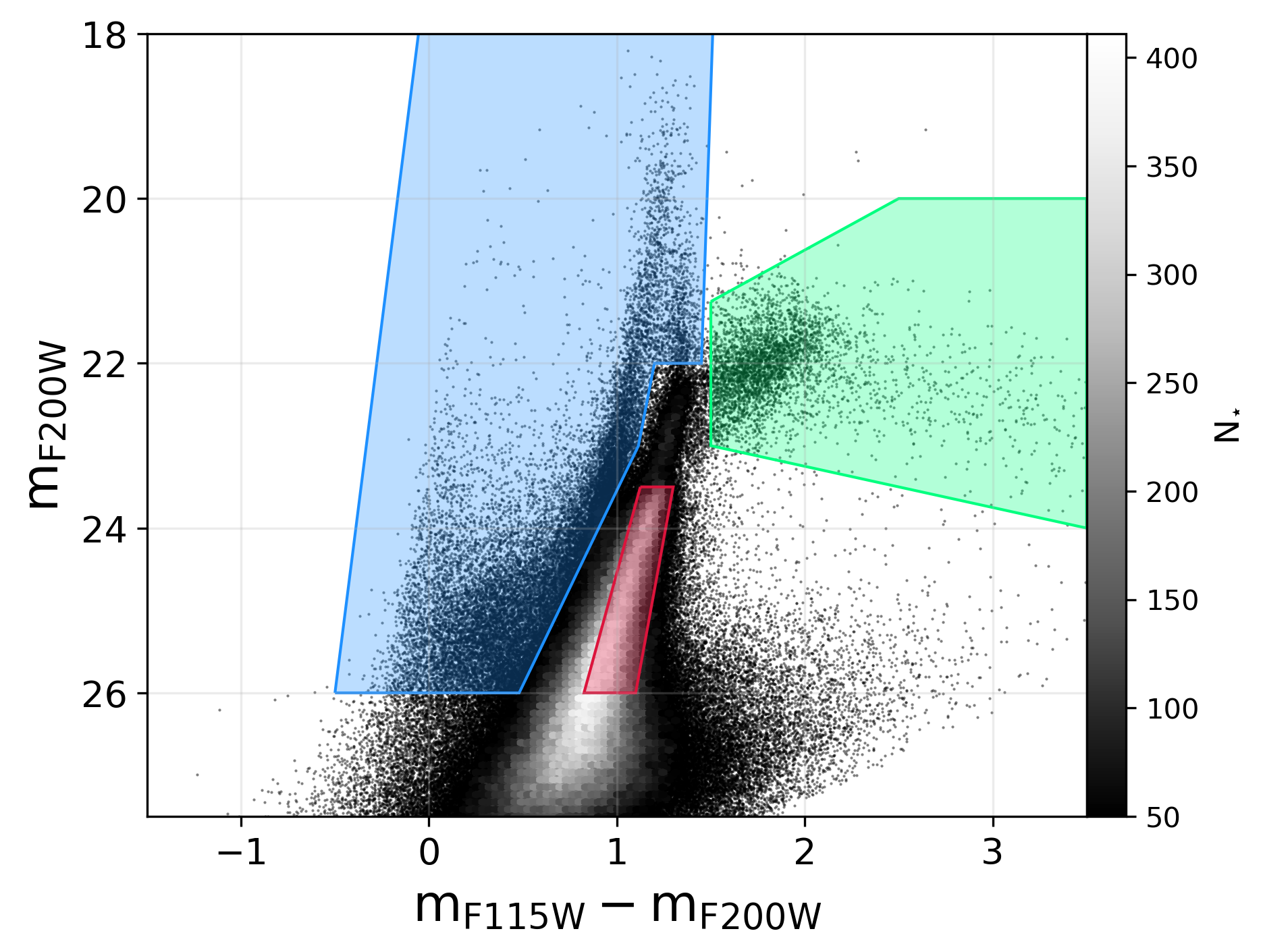}
    \caption{CMD of NGC~4485 (i.e., tile 1, 2, 3, 4, and 5) showing our selection of three different stellar populations, tracing different epochs in the life of the galaxy. MS, BL, RSG, and O-rich AGB stars younger than $200$ Myr in blue. C-rich AGBs between $\sim 200$ Myr and  $\sim 1$ Gyr in green, and RGB stars older than $\sim 1$ Gyr, and potentially as old as $\sim 13$ Gyr, in red.}
    \label{fig:8}
\end{figure}

To investigate the interaction history between NGC 4485 and NGC 4490, we isolate and select different regions in the CMD of both dwarf galaxies (see Section \ref{sec:Color-Magnitude Diagrams}), tracing different stellar phases and age ranges, to study their spatial distribution across our field of view. In particular, we identify three distinct CMD regions, with color and magnitude limits guided by careful comparisons between the CMDs and PARSEC-COLIBRI stellar isochrones (see Figure \ref{fig:6}). Figure \ref{fig:8} illustrates our selections of stars for NGC~4485's CMD, which includes tile 1, 2, 3, 4, and 5. Similar methodologies and selections have been applied to NGC~4490's CMD including the remaining tiles.

The shaded area in blue highlights young MS, RSG, as well as more evolved BL stars and O-rich AGBs. The strict magnitude limit at $\mathrm{m_{F200W}} = 26$ mag is imposed to minimize the influence of photometric errors leading to population mixing and incompleteness effects in our analysis. Additionally, we exclude redder stars to avoid contamination from older AGB and RGB stars. This selection corresponds to an age range $\lesssim 200$ Myr. Stars in the bright horizontal-red 'tail', shaded in green, are mostly C-rich AGB stars, tracing intermediate epochs, spanning from $\sim 200$ Myr to $\sim 1$ Gyr. We note that, due to the known uncertainties affecting carbon-rich AGB models \citep{Ventura2018}, the age limits associated with this selection are somewhat uncertain. Finally, the region shaded in red at $\mathrm{m_{F115W}} - \mathrm{m_{F200W}} \sim 1.25$ mag and $23.8$ mag $\lesssim \mathrm{m_{F200W}} \lesssim 26$ mag is dominated by low-mass RGB stars older than $\sim 1$ Gyr, and potentially as old as $\sim 13$ Gyr. Again, the $\mathrm{m_{F200W}} < 26$ mag limit is imposed to mitigate undesired effects due to large photometric errors affecting fainter stars and severe incompleteness that might weaken our conclusions.
\begin{figure*}[thbp!]
    \centering
    \includegraphics[width=1\textwidth]{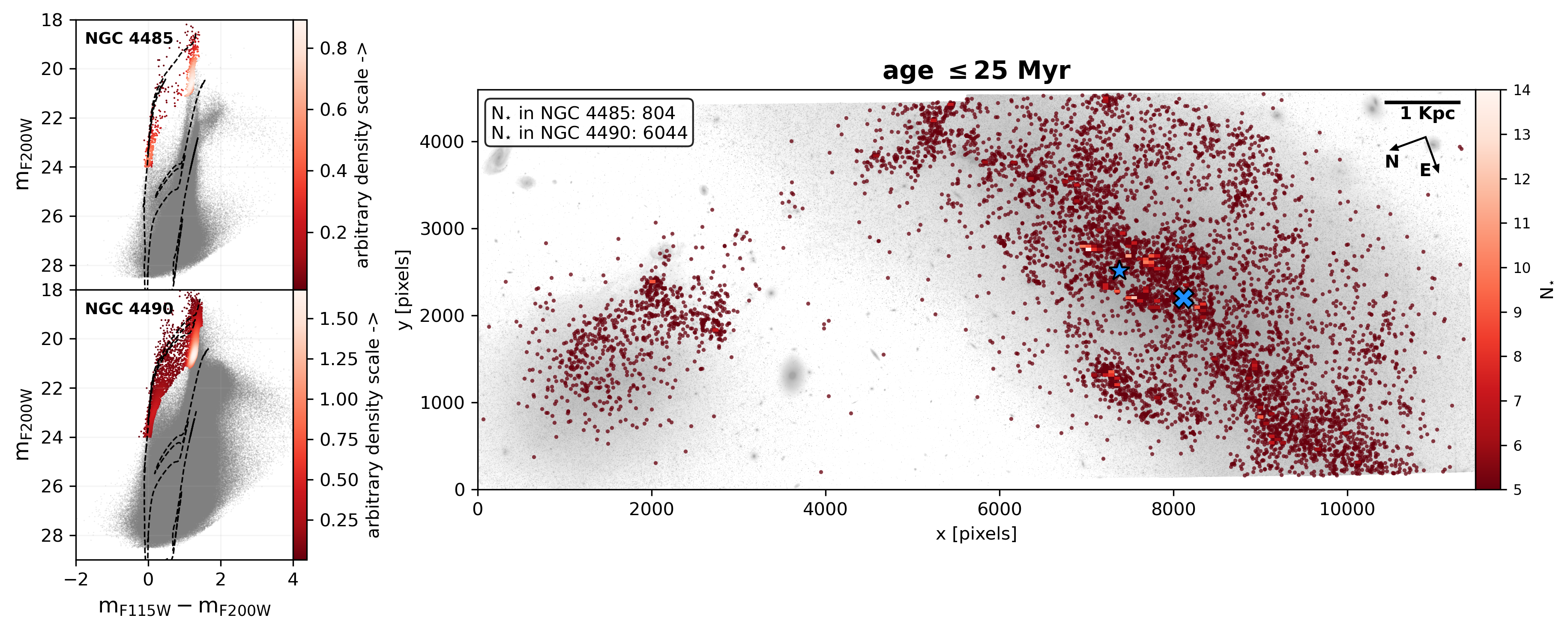}
    \includegraphics[width=1\textwidth]{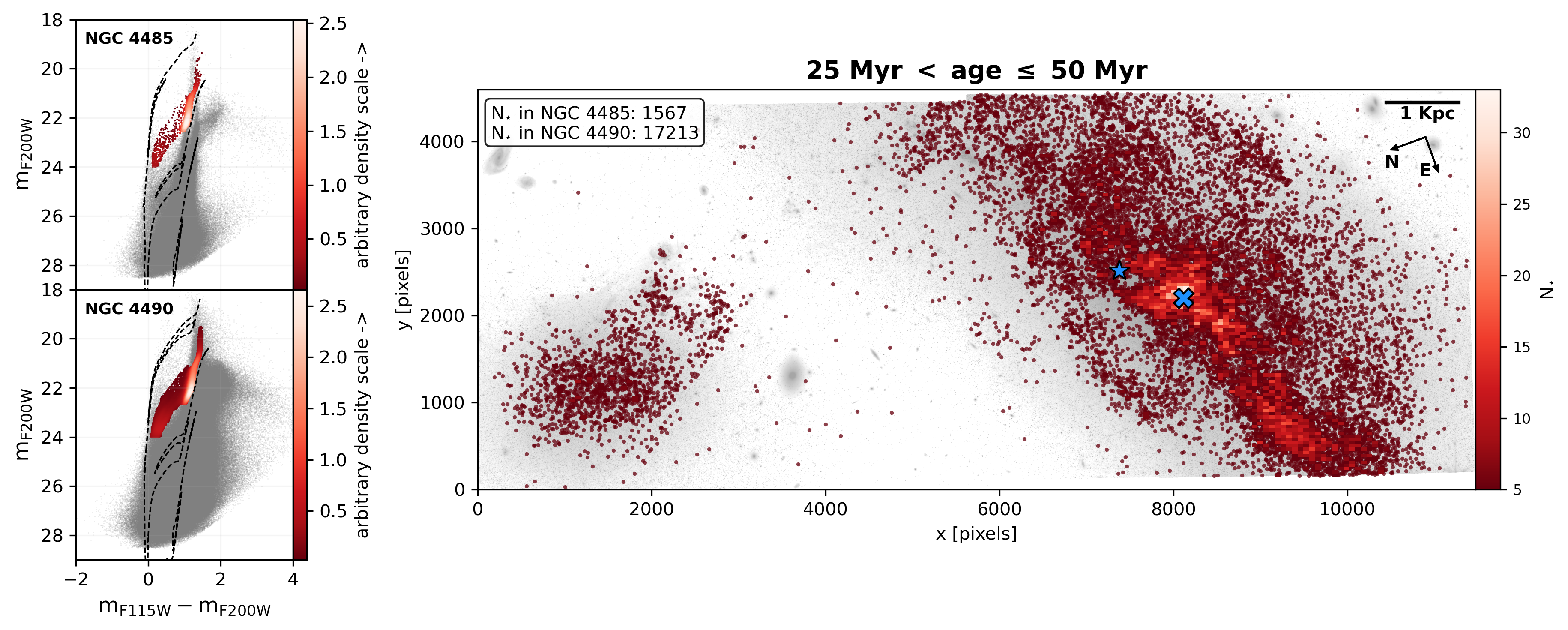}
    \includegraphics[width=1\textwidth]{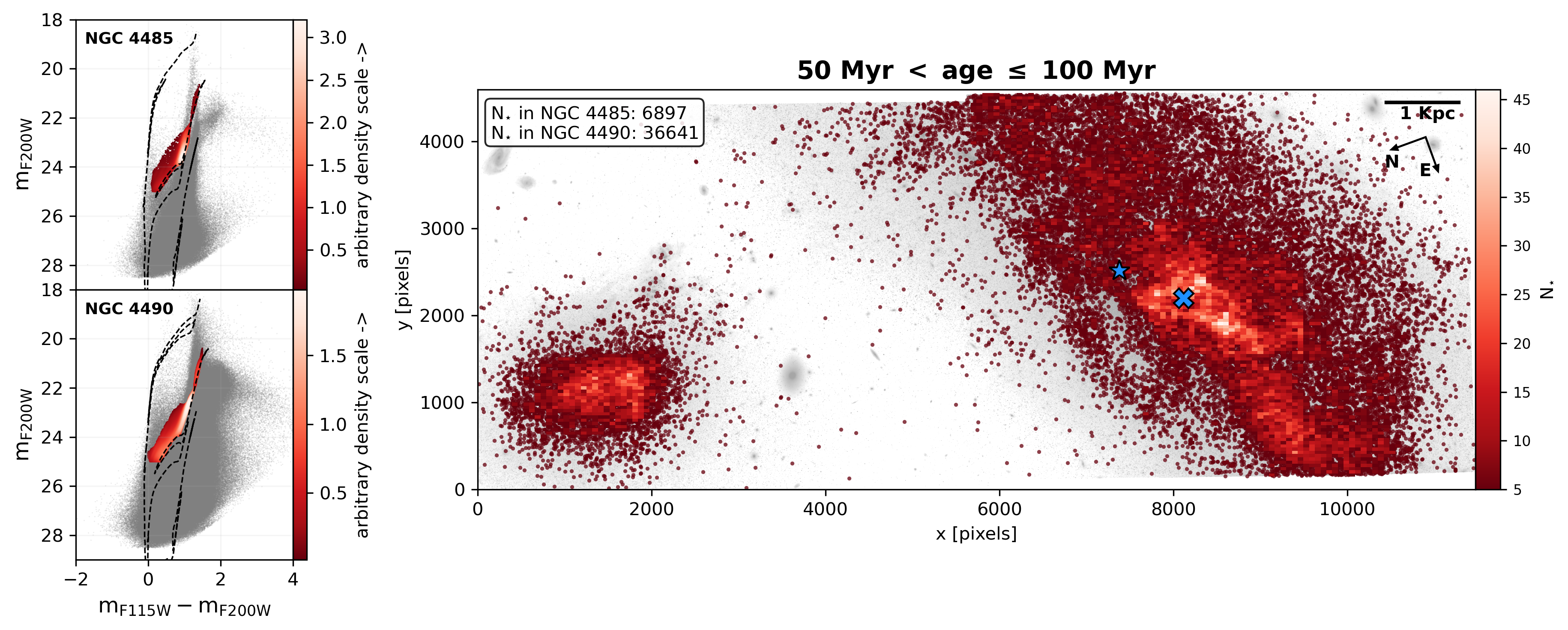}
    \caption{CMDs and spatial distributions of NGC~4485/NGC~4490 stars in different age intervals: age $\leq 25$ Myr, $25$ Myr $<$ age $\leq 50$ Myr, and $50$ Myr $<$ age $\leq 100$ Myr. Left panels: NGC~4485 (upper panel) and NGC~4490 (lower panel) CMDs, with the stars in the selected age intervals marked in red, and with the PARSEC-COLIBRI $10$ Myr, $100$ Myr, and $1$ Gyr isochrones in dashed-black lines. Right panels: Spatial distribution of the stars in the selected age intervals, overlaid on the F200W reference image. In the high-density regions, the data are binned and color coded according to the number of points (see the color-bar on the right hand-side of each panel). The cross and star symbols mark the position of the optical and infrared nuclei, as reported by \citet{Lawrence2020}. The number of stars in the selected age interval for both NGC~4485 and NGC~4490 is reported in the upper left.}
    \label{fig:9}
\end{figure*}

\begin{figure*}[thbp!]
    \centering
    \includegraphics[width=1\textwidth]{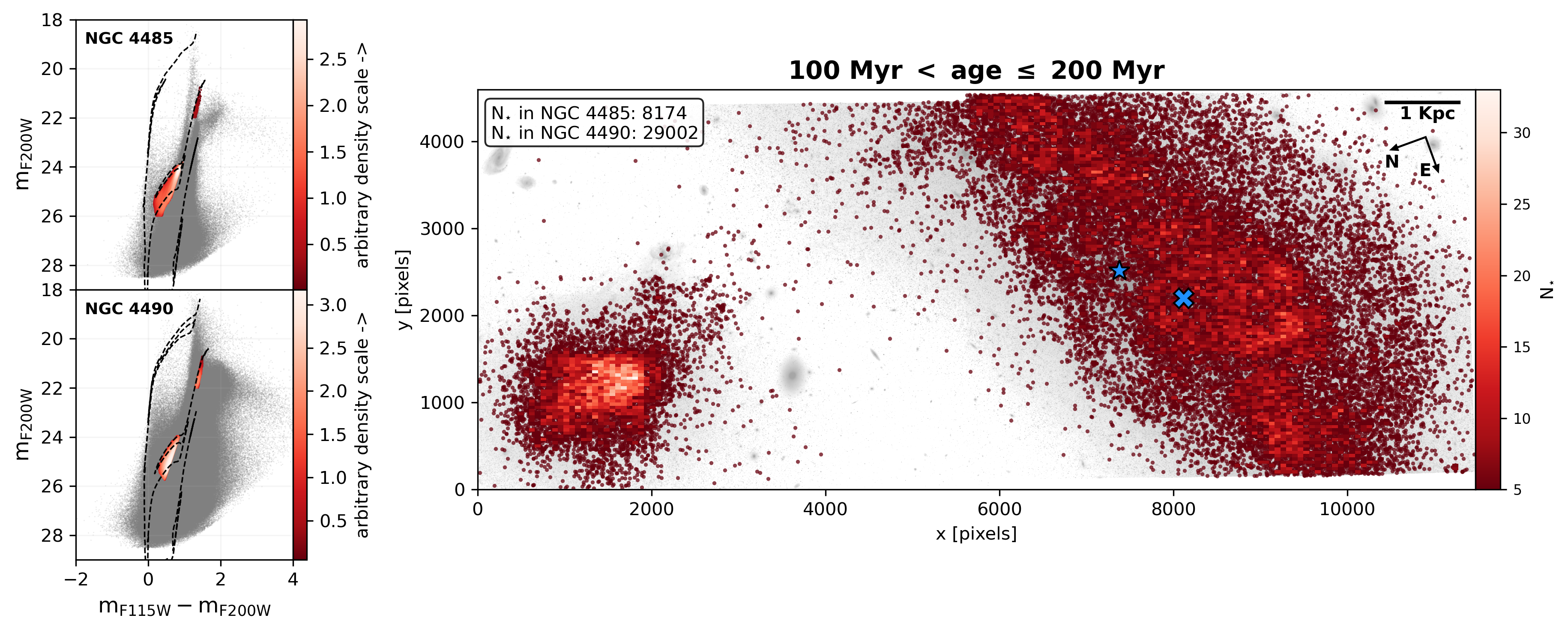}
    \includegraphics[width=1\textwidth]{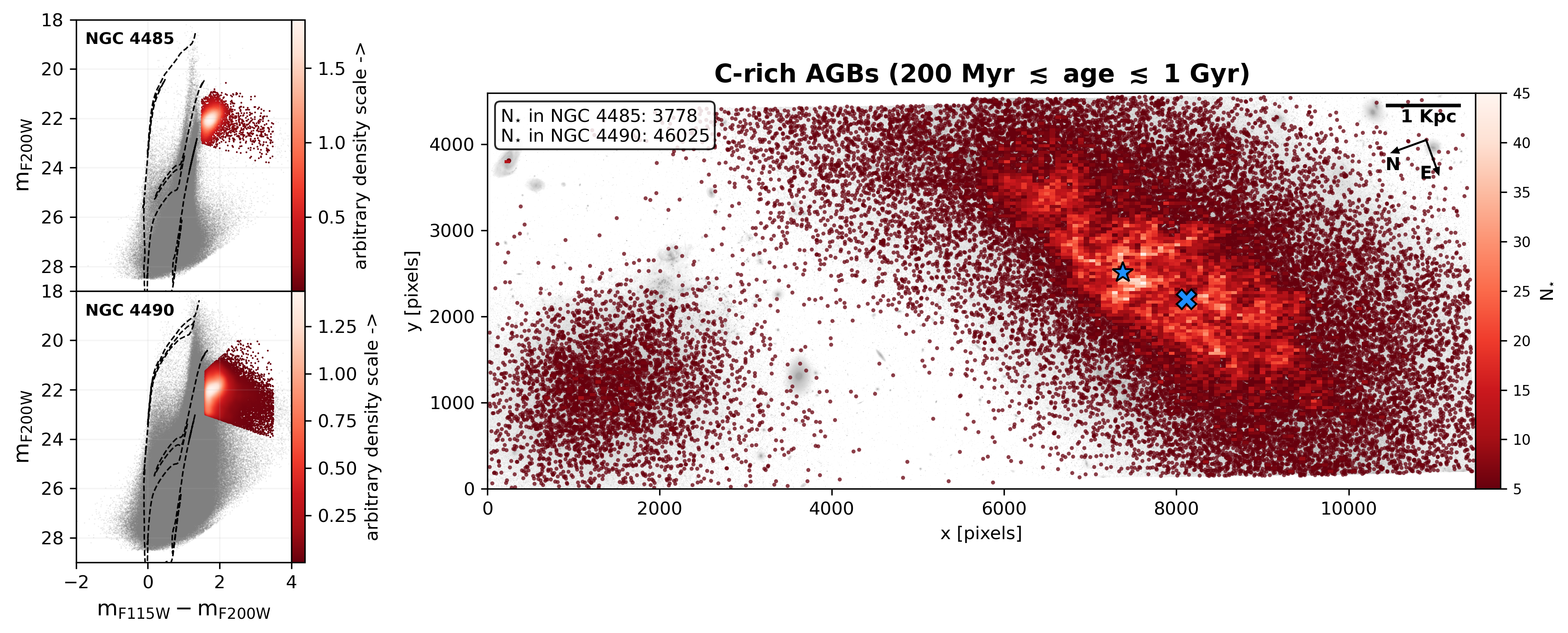}
     \includegraphics[width=1\textwidth]{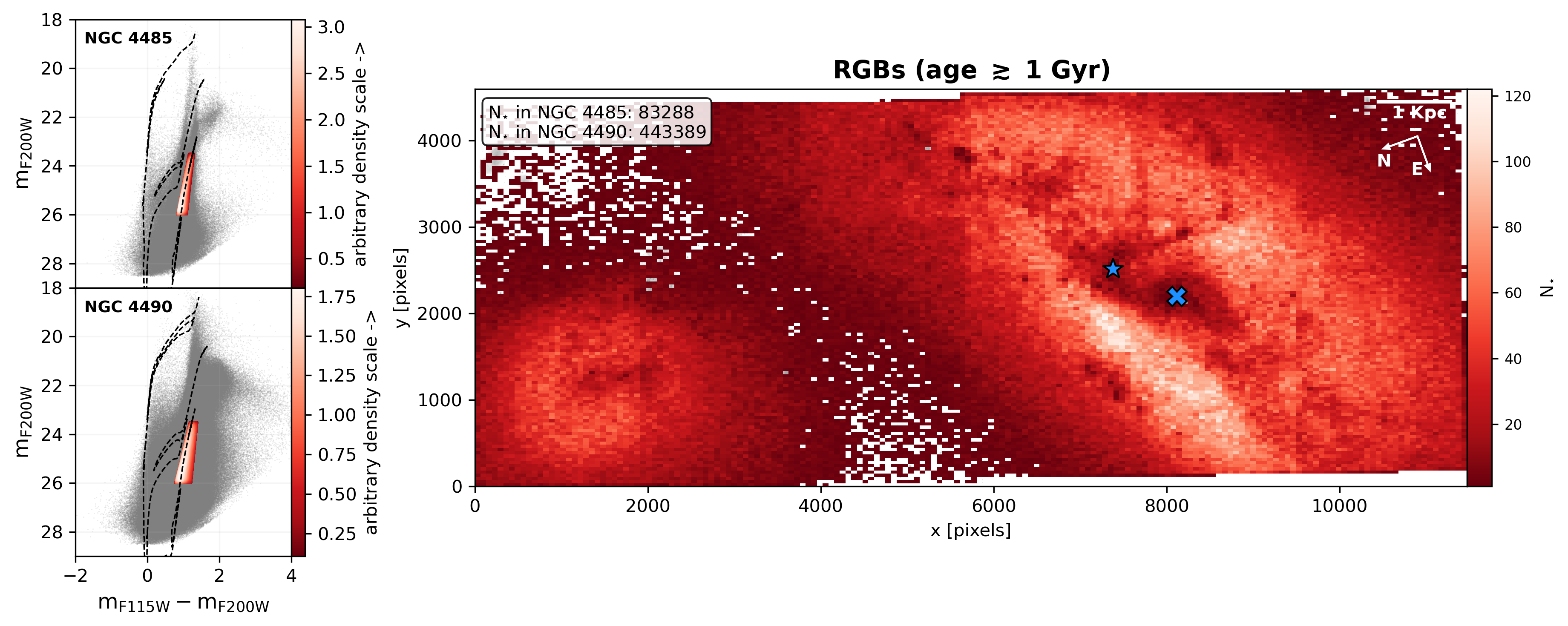}
    \caption{Same as Figure \ref{fig:9}, but for selected bona-fide stars with ages between $\sim 100$ Myr and $\sim 200$ Myr, intermediate age C-rich AGBs between $200$ Myr and $1$ Gyr, and RGB stars older than $1$ Gyr.}
    \label{fig:10}
\end{figure*}

In addition, we further divide our youngest stellar population selection (the one in blue in Figure \ref{fig:8}) into four smaller age bins: younger than $25$ Myr, from $25$ Myr to $ 50$ Myr, from $50$ Myr to $ 100$ Myr, and from $100$ Myr to $200$ Myr. The choice of this particular age binning size is driven by the need to have enough time sensitivity to best study the most recent star formation history, but without compromising our results due to small-number statistic or inherent population mixing due to uncertainties from photometric errors. To accomplish this we follow a procedure similar to what is described in \citet{Correnti2025b}. We first generated two well-populated synthetic CMDs, assuming PARSEC-COLIBRI stellar models, a Kroupa Initial Mass Function (IMF, \citealt{Kroupa2001}), a constant star-formation rate (SFR) in the last $800$ Myr, a distance modulus of $\mathrm{(m-M)_{0}} = 29.70$, a reddening of $\mathrm{E(B-V)} = 0.15$, and $\mathrm{[Fe/H] = -0.6}$ and $\mathrm{[Fe/H] = -0.3}$ for NGC~4485 and NGC~4490, respectively. Then, after convolving the `clean' synthetic CMDs with the photometric errors and completeness characteristic of each galaxy, we are ready to associate an age to each observed source. For each observed star, we selected all synthetic stars whose positions in the CMD lie within a radius of $0.1$ mag and computed the mean and standard deviation of their age distribution, which are then assigned to the observed star. If less than five synthetic stars are found within this radius, we progressively increase the radius in steps of $0.05$ mag until a minimum of five stars is reached. Finally, each observed star is assigned to its respective age bin.

Figures \ref{fig:9} and \ref{fig:10} show the spatial distribution of the stars belonging to the different age selections described above, overlaid on the F200W NIRCam image. In the left panels we show the total CMD of NGC~4485 and NGC~4490 in gray, with the selected stars belonging to that particular age interval in red. To help the reader, we also plot three isochrones (black-dashed lines), with ages of $10$ Myr, $100$ Myr, and $1$ Gyr, respectively. In the right panel, we binned the distribution of stars and plot the 2D-histogram color-coded according to the number of stars that fall in each cell (going from dark red to white, see the color-bar on the right-hand side). In the upper-left corner we reported the total number of selected stars in each galaxy. The blue cross and star symbols mark the location of the optical and dust-shrouded IR nuclei of NGC~4490 found by \citet{Lawrence2020}, respectively. 

\begin{figure*}[thbp!]
    \centering
    \includegraphics[width=1\textwidth]{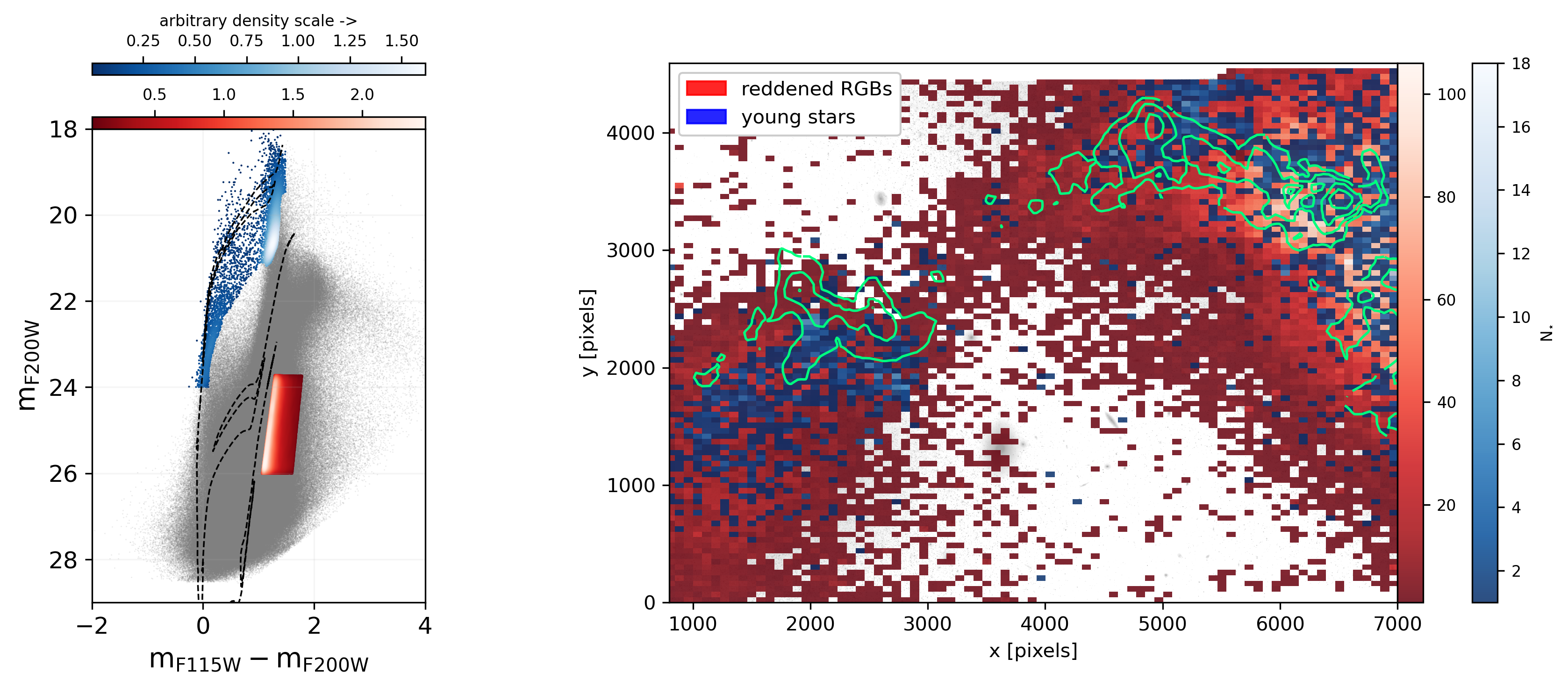}
    \caption{Left panel: CMD of the entire NIRCam FoV, with the selected populations of young stars (in blue) and reddened RGBs (in red). Right panel: comparison between the spatial distribution in the bridge region of reddened RGB stars (in red), and our selection of bona-fide young stars (age $\leq 25$ Myr, see top panel of Figure \ref{fig:9}) in blue. The data are color-coded according to their number density (see colorbars on the right hand-side). HI contours in green at $[4, 6, 8, 10] \times 10^{21} \, \mathrm{cm}^{-2}$ from a VLA B-configuration ($7''$ beam) pilot program (Cannon et al., priv. comm.) are also shown.}
    \label{fig:11}
\end{figure*}

Looking in more detail at the spatial distribution of our selected bona fide stars younger than $25$ Myr (upper panel of Figure \ref{fig:9}), we observe a distinct tidal tail extending from NGC~4485. This tail is connected to NGC~4485's center and is oriented toward NGC~4490 in the southwest direction, where it terminates in two compact and highly active star-forming regions at x $\sim 2000$ and y $\sim 2000$. In contrast, the young stellar population in NGC 4490 exhibits a more complex morphology. We identify a star-forming disk extending from the northwest to the southeast, as well as more prominent clumps of young stars in the spiral arms, particularly on the disk's northeast side at x $\sim 7000$ and y $\sim 1000$. At x $\sim 5000$ and y $\sim 4000$, a star forming complex extends toward NGC~4485 and completes the bridge structure.

Moving to the next age bin, i.e. $25$ Myr $\lesssim$ age $\lesssim 50$ Myr (middle panel of Figure \ref{fig:9}), we can still distinguish a well populated final section of NGC~4485's tidal tail, but complemented by a relative higher number of stars present in the main body of the dwarf galaxy. For NGC~4490, we observe a strong over-density in the position of the optical nucleus (marked by the blue cross, \citealt{Lawrence2020}), and a pronounced tail on the East side of the disk at x $\sim 9000$  and y $\sim 500$. Stars in the age interval between $50 \mbox{--} 100$ Myr (lower panel of Figure \ref{fig:9}) in NGC~4485 show a more concentrated distribution around the center of the galaxy, with an even lower ratio between the number of stars forming the bridge vs. the core. In NGC~4490, these stars are distributed in a similar fashion compared to the previous age bin, except for the appearance of what seems like a second high-density peak positioned to the East of the optical nucleus, around x $\sim 8500$ and y $\sim 1500$.

The top panel of Figure \ref{fig:10} shows the spatial distribution of our selected stars from $100$ to $200$ Myr. These stars trace the time frame in which \cite{Pearson2018} predicted the latest close pericenter passage between the two dwarfs to happen. NGC~4485 displays a high concentration of stars in its core with respect to the star-forming regions in the bridge. This finding seems to confirm that a strong starburst occurred in the center of NGC~4485, probably triggered by the compression of the gas due to the interaction with NGC~4490 \citep{Elmegreen1998}. This passage is also likely responsible for the stripping of the HI gas that later collapsed to form the bright HII regions in the bridge. On the other hand, NGC~4490 shows all the features already present in the previous age intervals, with no particular over-densities in both nuclei. We warn the reader that this lack of peaks in the center of NGC~4490 might be due to the intrinsic lower completeness in this region. Nevertheless, NGC~4490's disk, spiral arms, and tidal tails are still visible, suggesting that the interaction with the companion has triggered spatially diffuse star formation episodes across the galaxy. 

The middle panel of Figure \ref{fig:9} shows the spatial distribution of our selection of intermediate age ($200$ Myr $\lesssim$ age $\lesssim 1$ Gyr) C-rich AGB stars, forming the horizontal red `tail' in the CMDs. As expected, their distribution in both dwarf galaxies looks less clumped compared to younger stars, and more uniform across the bodies of the two dwarf galaxies. Interestingly, we report that NGC~4490 seems to show a peak of stars close to the location of the infrared nucleus reported by \citet{Lawrence2020}. Moreover, in the region between the two dwarfs, we recovered some stars bridging between the two halos. 

The bottom panel of Figure \ref{fig:10} shows the spatial distribution of a sample of old (age $\gtrsim 1$ Gyr) RGB stars. The intrinsic faintness of these stars makes them prone to suffer from completeness issues at this distance. For this reason, we restricted our samples for both galaxies from $\mathrm{m_{F200W}} = 23.8$ mag (the location of the tip) to $\mathrm{m_{F200W}} = 26$ mag. Despite this precaution, most `holes' in the spatial distribution of this sample, especially closer to the centers of the galaxies and bright star forming regions are most likely due to a combination of incompleteness and possibly extinction from dust lanes (see red channel in Figure \ref{fig:2}). The long life of these stars makes them subject to the effects of orbital mixing \citep{Ibata1997,Helmi1999,Cooper2010,Belokurov2018}, thus they show a broad and uniform distribution, occupying most of the field of view. We report, somewhat surprisingly, that the old stellar halos of the two dwarf galaxies seem not to have been strongly perturbed despite the close gravitational interaction between the two bodies \citep[see e.g.][]{Pascale2024,Sacchi2024}. Nevertheless, this is most likely due to our restricted field of view (see Figure \ref{fig:1}), since faint and extended tidal features have been observed in deep ground-based imaging \citep[see e.g.,][]{ElmegreenD1998,Smith2010,Pearson2018,Lawrence2020,Mandal2024}.

Finally, in Figure \ref{fig:11}, we compare the spatial distribution of a selection of RGB stars that are redder than the average color predicted by stellar models, along with our previous selection of RGB stars (see left panel), to our sample of stars younger than 25 Myr (see top panel of Figure \ref{fig:9}). We focus in particular on the bridge region, where the RGB stars appear to outline a fully connected bridge between the two dwarf galaxies, around $x \sim 4000$ and $y \sim 3000$, aligned with the positions of the youngest stars (shown in blue) and HI contours from a VLA B configuration pilot program (Cannon et al., priv. comm). These RGB stars are likely redder due to the presence of dust mixed with the gas. Therefore, this population can be used to trace regions of higher gas density and extinction. This result reinforces the idea that a gas/dust lane connects the two dwarf galaxies, likely created during their most recent close encounter through tidal and ram pressure stripping, and currently sustaining the ongoing star formation observed in these regions \citep{Mandal2024}.

\section{Summary \& Conclusions} \label{sec:summary and conclusions}
In this paper, we presented new JWST/NIRCam observations of the closest known interacting dwarf galaxy pair, NGC 4485/NGC 4490, obtained as part of the Cycle 1 FEAST program (Adamo et al., in prep). This system offers a unique laboratory for studying galaxy interactions at the smallest scales, providing valuable insights into how such processes may shape galaxy formation and evolution at high redshift. Our main goal was to probe, for the first time, the resolved stellar populations of this remarkable system in the near-infrared, leveraging JWST’s exquisite capabilities. Here we summarize our main findings:

\begin{itemize}
    
    \item  The NIRCam data (see Figure \ref{fig:2}) confirms the presence of a young stellar bridge connecting the two dwarf galaxies, composed of bright HII regions that trace a dense gas bridge fueling the active star formation. This bridge is likely the result of tidal and/or ram pressure stripping that has occurred during the last close encounter between the two companions \citep{Clemens2000,Pearson2018,Mandal2024}. 

    \item The NIR CMDs of the galaxies host an impressive variety of stellar populations from tile to tile, spanning a substantial range in age and metallicity. Both galaxies show  prominent populations of young ($\lesssim 200$ Myr) upper-MS, BL, and O-rich AGB stars. In particular, we report the presence of well populated RSG branches reaching in some cases up to $\mathrm{m_{F200W}} \sim 18$ mag. We also detect a substantial population of intermediate-age C-rich AGB stars, occupying a well defined red `tail' in the CMDs. We recover a robust RGB sequence populated by low-mass stars, older than 1 Gyr.

    \item Tiles 5 and 9 (which cover the bridge region), along with tile 10 (a star-forming region on the northeastern side of NGC~4490), show a clear gap in their RSG sequences at $\mathrm{m_{F200W}} \sim 23$, followed by a prominently populated brighter section. This morphology can only be explained by an intense burst of star formation between $\sim 5$ and $30$ Myr ago. Moreover,  we found a significant difference in the average color ($\sim 0.2$ mag) of the RSG sequences of tile 5 and 9. This difference in color translates to a factor of two difference in the metal content of the respective young stellar populations (assuming a solar abundance pattern). This result strongly supports the scenario in which metal-poor gas stripped from NGC 4485 was mixed with the more metal-rich gas of NGC 4490, before forming the young population of stars we observe in the bridge region and around the main body of NGC 4490.

    \item In NGC~4485, we identify an over-density of blue He-burning stars around $\mathrm{m_{F115W}} - \mathrm{m_{F200W}} \sim 0.5$ mag and $\mathrm{m_{F200W}} \sim 25$ mag. The properties of this stellar evolutionary phase allow us to precisely date the associated burst of star formation to approximately $100\mbox{--}200$ Myr ago. Our findings are in very good agreement with the independent predictions by \citet{Pearson2018}, based on N-body simulations, which place the system's last pericenter passage at around $230$ Myr ago.

    \item We compared the spatial distributions of a sample of bona-fide stellar populations, selected from the observed CMDs using synthetic CMDs generated from PARSEC-COLIBRI stellar models, tracing different epochs spanning the entire history of the system. The spatial distribution of the youngest stars ($\leq 25$ Myr) shows a tidal tail extending from the center of NGC 4485 connecting to NGC 4490 disk through the bridge. NGC 4490 shows also compact star forming regions associated with its spiral arms that originates from the IR nucleus. The population tracing the age range $100\mbox{--}200$ Myr shows a significant over-density of stars in the core of NGC 4485 with respect to the bridge region. This result supports the hypothesis of an intense central starburst in NGC 4485, likely initiated by gas compression resulting from its recent close interaction with NGC 4490. Finally, the stellar populations tracing older epochs (i.e. AGBs and RGBs) show a more uniform distribution, with no particular evidence of perturbation.

    \item The spatial distribution of a selection of reddened RGB stars closely traces the location of the young star forming regions identifying a continuous bridge connecting the two interacting galaxies (see Figure \ref{fig:11}). These stars appear redder due to the presence of dust within the HI gas. This result suggests the presence of a dense gas/dust lane connecting the two dwarf galaxies, as confirmed by independent VLA B configuration observations (Cannon et al., priv. comm).

\end{itemize}

The study of the resolved stellar populations presented in this paper, combined with previous N-body simulations \citep{Pearson2018} and upcoming investigations of the ISM properties (Duarte Puertas et al., in prep), adds upon and improves our picture of the interaction history of the NGC 4485/NGC 4490 dwarf galaxy system. This work represents a first step toward a more detailed analysis of our JWST data, which will include deriving the system’s spatially resolved star formation history through CMD modeling (Bortolini et al., in prep). Our results, together with other recent studies of resolved stellar populations in Local Volume galaxies \citep[see, among others,][]{Weisz2023,McQuinn2024,Habel2024,Nally2024,Bortolini2024b,Correnti2025b}, demonstrate the remarkable capabilities of JWST, far exceeding those of previous infrared telescopes and providing unprecedented insights into the star formation processes of nearby dwarf galaxies.

\section*{Acknowledgments}
The authors wish to thank the anonymous reviewer for their insightful comments, which helped improve the manuscript. AA acknowledges support from Vetenskapsr\aa det 2021-05559. A.A and A.P. acknowledge support from the Swedish National Space Agency (SNSA) through the grant 2021- 00108. A.A. and H.F.V. acknowledges support from (SNSA) 2023-00260. KG is supported by the Australian Research Council through the Discovery Early Career Researcher Award (DECRA) Fellowship (project number DE220100766) funded by the Australian Government. KG is supported by the Australian Research Council Centre of Excellence for All Sky Astrophysics in 3 Dimensions (ASTRO~3D), through project number CE170100013. This work is based in part on observations made with the NASA/ESA/CSA James Webb Space Telescope, which is operated by the Association of Universities for Research in Astronomy, Inc., under NASA contract NAS 5-03127.
These observations are associated with program \# 1783. Support for program \# 1783 was provided by NASA through a grant from the Space Telescope Science Institute,
which is operated by the Association of Universities for Research in Astronomy, Inc., under NASA contract NAS 5-03127. MM acknowledges financial support through grants PRIN-MIUR 2020SKSTHZ, the INAF GO Grant 2022 “The revolution is around the corner: JWST will probe globular cluster precursors and Population III stellar clusters at cosmic dawn,” and by the European Union – NextGenerationEU within PRIN 2022 project n.20229YBSAN - "Globular clusters in cosmological simulations and lensed fields: from their birth to the present epoch. SDP acknowledges financial support from Ministerio de Economía y Competitividad under grants PID2022-136598NB-C32 ``Estallidos8'', PID2023-150178NB-I00, PID2023-149578NB-I00  financed by MICIU/EI/10.13039/501100011033/ and FEDER, UE and from Junta de Andalucía Excellence Project P18-FR-2664 and FQM108.

\section*{Data Availability}
The data supporting the findings of this study are openly available at the following MAST DOI: \url{http://dx.doi.org/10.17909/ybh0-r149}. All our data products are available at MAST as a High Level Science Product via \url{https://doi.org/10.17909/6dc1-9h53} and DOI: 10.17909/6dc1-9h53 and on the FEAST webpage \url{https://feast-survey.github.io/}.

%

\vspace{5mm}
\facilities{JWST/NIRCam.}


\software{{\tt DOLPHOT} \citep[\url{http://americano.dolphinsim.com/dolphot/};][]{Dolphin2000, Dolphin2016}, {\tt Astropy} \citep{astropy:2013,astropy:2018,astropy:2022}, {\tt Numpy} \citep{Numpy2020}, {\tt Pandas} \citep{Pandas2010}, {\tt SciPy} \citep{Scipy2020}, {\tt Matplotlib} \citep{Matplotlib2007}, {\tt SAOImageDS9} (developed by the Smithsonian Astrophysical Observatory 2000, \citealt{Joye2003}), {\tt TOPCAT} \citep{Topcat2005}.
          }



\appendix

\section{Analysis of Color-Magnitude Diagrams for the Remaining Tiles} \label{app:Analysis of the remaining CMDs}
\begin{figure*}[thbp!]
    \centering
    \includegraphics[width=1\textwidth]{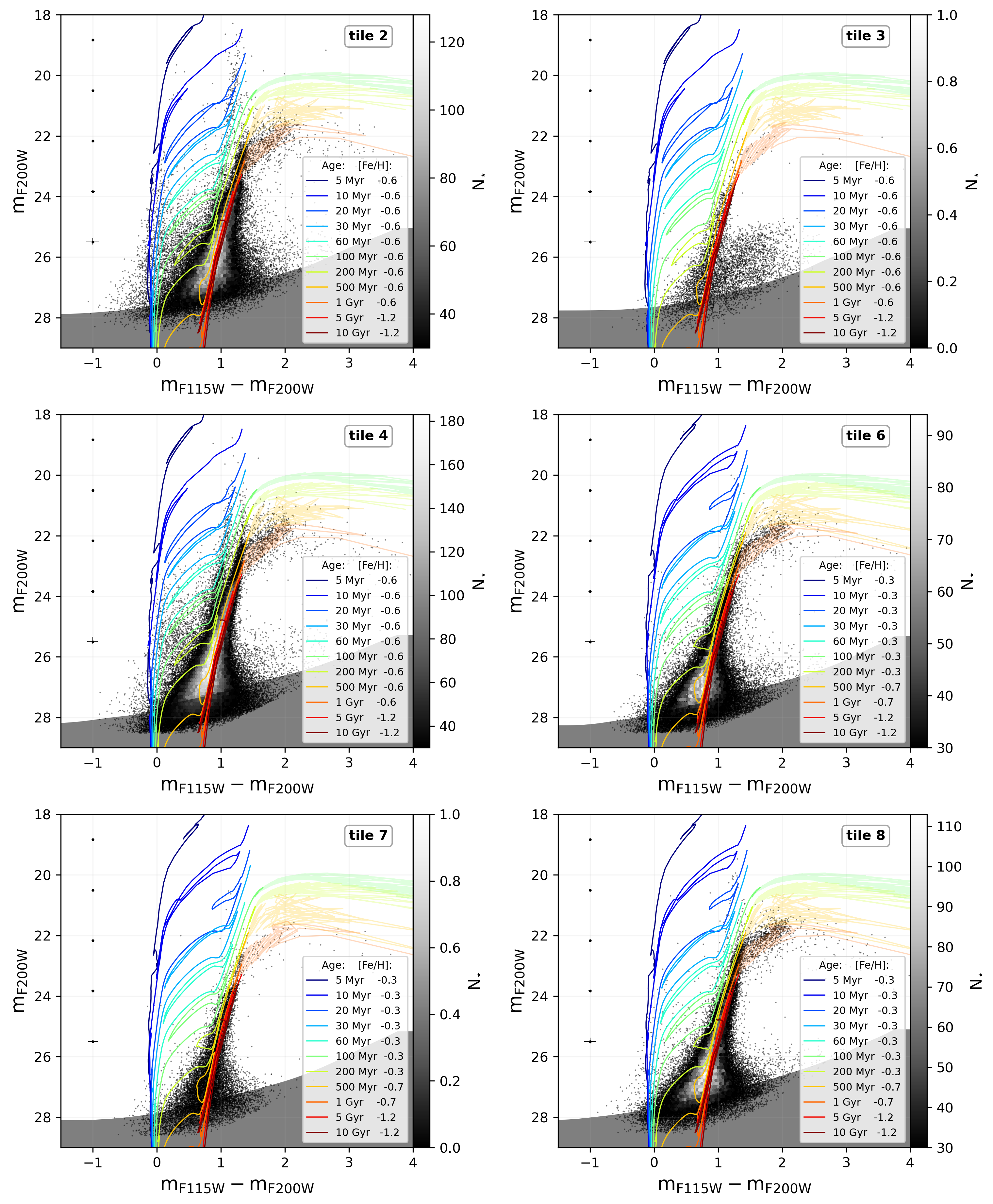}
    \caption{CMDs of tile 2, 3, 4, 6, 7, and 8. See Figure \ref{fig:6}.}
    \label{fig:appx_1}
\end{figure*}
\begin{figure*}[thbp!]
    \centering
    \includegraphics[width=1\textwidth]{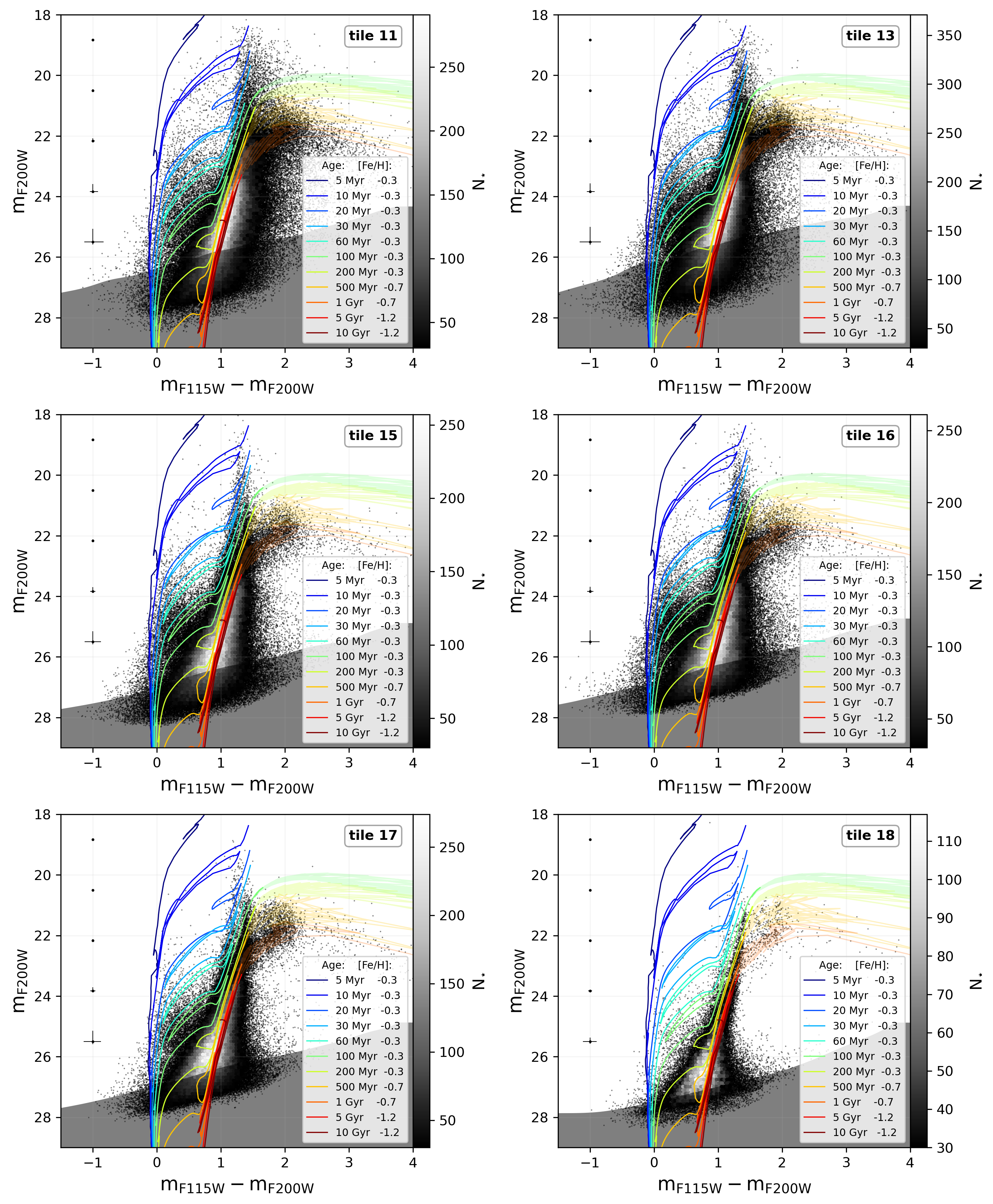}
    \caption{CMDs of tile 11, 13, 15, 16, 17, and 18. See Figure \ref{fig:6}.}
    \label{fig:appx_2}
\end{figure*}
Figures \ref{fig:appx_1} and \ref{fig:appx_2} show the CMDs for the remaining tiles not shown in the main text. PARSEC-COLIBRI stellar isochrones are over-plotted and color-coded by their age and metallicity as in Figure \ref{fig:6}. Photometric errors as a function of magnitude (error bars on the left-hand side) and the $50\%$ completeness limits (dark-shaded area) are also reported. Tile 2, which covers part of the main body of NGC~4485, displays many similarities with tile 5, including a prominent population of bright RSG stars, with a gap around $\mathrm{m_{F200W}} \sim 23$ mag. Moreover, as in tile 5, it does not show any trace of the red O-rich AGB `finger' parallel to the RSG sequence, which is instead clearly visible in the other tiles covering the main body of NGC~4485, namely tiles 1 and 4. Interestingly, tile 4 presents a well-defined RSG sequence, but without the gap seen in tile 5. Tiles 3 and 7 are probably the simplest, displaying clear RGBs dominated by old ($\gtrsim 1$ Gyr) stars, with no signs of recent star formation. The very faint ($\mathrm{m_{F200W}} \sim 26$ mag) and red ($\mathrm{m_{F115W}} - \mathrm{m_{F200W}} \sim 1.5$ mag) sources that can be seen in tile 3 consist mostly of background galaxies. Indeed, due to the intrinsic low number of stars detected in this tile, the sharpness and crowding sigma clipping that was applied to the other tiles is too restrictive. We chose not to modify the cuts since we do not encounter the same problem in more populated tiles and given that we did not use that particular region of the CMD in any of our analysis. Tiles 6 and 8 cover the outer portion of the tidal bridge between the two dwarf galaxies. Both diagrams show a prominent old RGB, topped by a well-populated red `tail' of C-rich AGB stars. In both diagrams, only sparse signs of recent star formation are visible, with a few stars found in the upper MS and BL phases. 

Most of the central tiles of NGC~4490, i.e. tile 11, 13, 15, 16, and 17, show the same main features discussed for NGC~4485's CMDs (even if affected in some case by a shallower and more uncertain photometry due to the more intense crowding conditions). A very bright RSG sequence, a well populated BL sequence corresponding to the isochrones of ages between $\sim 100$ and $200$ Myr, always complemented by a O-rich red `finger' to the right of the RSG sequence. A pronounced red horizontal tail of TPAGBs, and a dominant old RGB. As expected, tile 18, which covers part of the outskirts of NGC~4490, is predominantly populated by old RGB stars. However, it also exhibits a sparsely populated group of BL stars around $\mathrm{m_{F200W}} \sim 25$ mag, resembling the $100 \mbox{--} 200$ Myr populations observed in other tiles of NGC~4490 and in NGC~4485. This suggests that traces of a recent pericenter passage between the two galaxies are also impressed in the low-surface brightness outskirts of the system.


\bibliography{biblio}{}
\bibliographystyle{aasjournal}



\end{document}